\documentclass[12pt]{article}
\pdfoutput=1
\usepackage{jheppub}
\usepackage{amsmath,amssymb,bbm}
\usepackage[font=small]{caption}
\usepackage[usenames,dvipsnames, table]{xcolor}
\usepackage{graphics, subcaption}
\usepackage{tikz}
\usepackage{multirow}
\usepackage{bm}
\usepackage[normalem]{ulem}
\usepackage{pifont}

\usepackage[framemethod=TikZ]{mdframed}
\mdfsetup{skipabove=0pt,skipbelow=0pt}
\mdfdefinestyle{boxed}{%
    linecolor=black,
    outerlinewidth=0.5pt,
    roundcorner=5pt,
    innertopmargin=\baselineskip,
    innerbottommargin=\baselineskip,
    innerrightmargin=20pt,
    innerleftmargin=20pt,
    backgroundcolor=gray!05!white
    }

\def\BigColSep{\setlength{\arraycolsep}{15pt}}

\newcommand{\bfp}{\mathbf{p}}

\begin{document}

\preprint{CERN-TH-2021-122}
\title{Natural Selection Rules:\\New Positivity Bounds for Massive Spinning Particles}

\author[a]{Joe Davighi,}
\author[a]{Scott Melville,}
\author[a, b]{and Tevong You}
\affiliation[a]{DAMTP, Center for Mathematical Sciences, University of Cambridge, CB3 0WA, UK}
\affiliation[b]{Theoretical Physics Department, CERN, CH-1211 Geneva 23, Switzerland}

\emailAdd{jed60@cam.ac.uk}
\emailAdd{scott.melville@damtp.cam.ac.uk}
\emailAdd{tevong.you@cern.ch}

\abstract{
We derive new effective field theory (EFT) positivity bounds on the elastic $2\to2$ scattering amplitudes of massive spinning particles from the standard UV properties of unitarity, causality, locality and Lorentz invariance. 
By bounding the $t$ derivatives of the amplitude (which can be represented as angular momentum matrix elements) in terms of the total ingoing helicity, we derive stronger unitarity bounds on the $s$- and $u$-channel branch cuts which determine the dispersion relation.
In contrast to previous positivity bounds, which relate the $t$-derivative to the forward-limit EFT amplitude with no $t$ derivatives, our bounds establish that the $t$-derivative alone must be strictly positive for sufficiently large helicities. 
Consequently, they provide stronger constraints beyond the forward limit and can be used to constrain dimension-6 interactions with a milder assumption about the high-energy growth of the UV amplitude.
}


\maketitle

\newpage
\section{Introduction}

The analytic properties of scattering amplitudes provide a rich connection between higher-dimensional operator coefficients in an effective field theory (EFT) and fundamental properties of its UV completion, particularly unitarity, causality and locality~\cite{Pham:1985cr, Ananthanarayan:1994hf, Pennington:1994kc, Adams:2006sv}. 
Rather than match EFT coefficients to a particular UV model, so-called ``positivity bounds" instead leverage such fundamental UV properties to constrain the EFT coefficient space.
In this work, we derive new and more general positivity bounds on EFTs for massive spinning particles, adding to the rapidly growing list of available bounds developed recently in \cite{Bellazzini:2016xrt,deRham:2017avq,deRham:2017zjm,Remmen:2020uze,Bellazzini:2020cot,Tolley:2020gtv,Caron-Huot:2020cmc, Sinha:2020win, Trott:2020ebl, Li:2021cjv, Arkani-Hamed:2020blm, Chiang:2021ziz}. 

It is surely no coincidence that this recent progress comes at a time when the lack of new physics near the weak scale challenges our understanding of EFT naturalness. 
This return to the foundations of QFT is reminiscent of the original analytic $S$-matrix programme~\cite{Chew, Eden}, which also began in an era when the contemporary understanding of QFT was being challenged. Ironically, back then this was due to an abundance of new resonances at the GeV scale; today, it is the lack of on-shell resonances at the TeV scale. Nevertheless, signatures of new physics may still arise from exploring the precision frontier, by detecting deviations from SM predictions. 
If such deviations are due to heavy new physics, then they are efficiently encoded as higher-dimensional operators in the Standard Model Effective Field Theory (SMEFT) framework, and positivity bounds provide a direct connection between these operator coefficients and general properties of the UV. 
Positivity bounds are an important theoretical prior restricting the EFT parameter space which could improve our estimation of these SMEFT parameters and therefore our ability to detect new physics in precision experiments. Conversely, experimental measurements probing the parameter space with the ``wrong" sign can test whether fundamental principles break down in the UV. 
%

The tantalising prospect of such a two-way bridge between the IR and the UV has led to a diverse range of phenomenological applications of positivity bounds. As mentioned before, LHC measurements are increasingly interpreted in terms of SMEFT coefficients 
(see e.g. Refs.~\cite{Ellis:2020unq, Ethier:2021bye} for the latest global fits) 
to which positivity bounds have recently been applied~\cite{Bellazzini:2017bkb, Bellazzini:2018paj,Zhang:2018shp,Bi:2019phv, Remmen:2019cyz, Englert:2019zmt, Remmen:2020vts, Bonnefoy:2020yee}.  
Other particle physics applications include chiral perturbation theory~\cite{Pham:1985cr,Ananthanarayan:1994hf,Pennington:1994kc} and its gauged siblings~\cite{Distler:2006if,Vecchi:2007na}. 
Positivity bounds have also been recently applied to a variety of EFTs relevant for cosmology, including the study of corrections to general relativity~\cite{Bellazzini:2015cra,Cheung:2016wjt,Camanho:2014apa,Gruzinov:2006ie}; the effective theories describing massive gravity~\cite{Cheung:2016yqr, Bonifacio:2016wcb, Bellazzini:2017fep, deRham:2017xox, deRham:2018qqo, Alberte:2019xfh, Alberte:2019zhd} and higher-spin states~\cite{Hinterbichler:2017qyt,Bonifacio:2018vzv, Bellazzini:2019bzh}; various scalar field theories~\cite{Nicolis:2009qm,Elvang:2012st,deRham:2017imi,Chandrasekaran:2018qmx,Herrero-Valea:2019hde}; Einstein--Maxwell theory and the Weak Gravity Conjecture~\cite{Cheung:2014ega,Cheung:2018cwt,Cheung:2019cwi,Bellazzini:2019xts,Charles:2019qqt}; and paired with observational constraints \cite{Melville:2019wyy, deRham:2021fpu}.
Any improvement in the constraining power of positivity bounds can therefore impact a wide range of areas throughout theoretical physics.

Following the influential study of Ref.~\cite{Adams:2006sv}, which developed bounds for the $2 \to 2$ scattering of scalar particles in the forward limit, positivity bounds have since been extended beyond the forward limit \cite{Nicolis:2009qm, deRham:2017avq}, and to include spinning particles \cite{Bellazzini:2016xrt, deRham:2017zjm}. Steps have also been taken beyond $2 \to 2$ processes \cite{Chandrasekaran:2018qmx}, to allow for spontaneous Lorentz breaking \cite{Baumann:2015nta, Grall:2020tqc, Grall:2021xxm, Aoki:2021ffc}, and to include gravitational effects~\cite{Alberte:2020bdz, Alberte:2020jsk, Tokuda:2020mlf, Herrero-Valea:2020wxz, Caron-Huot:2021rmr}.
Positivity bounds have recently been further strengthened by exploiting full crossing symmetry ({\em i.e.} equating the $s$-, $t$-, and $u$-channels) \cite{Tolley:2020gtv, Caron-Huot:2020cmc, Sinha:2020win}, moment theorems \cite{Bellazzini:2020cot}, and an emerging geometrical understanding of their underlying mathematical structure \cite{Arkani-Hamed:2020blm, Chiang:2021ziz}.
While all these bounds begin constraining operators at mass dimension-8 and higher, Ref.~\cite{Remmen:2020uze} recently pursued a complementary direction, in deriving positivity bounds on the dimension-6 interactions between two massless spinors by requiring a stronger convergence of the UV amplitude at high energies. 

However, despite these recent advances, it seems clear that the existing bounds are far from the full picture. Indeed, most explicit UV completions that we know of seem to populate only a small island in the region allowed by current positivity bounds \cite{Bern:2021ppb}. 
Improving on the existing positivity machinery, to find the strongest possible bounds on the space of low-energy EFTs, is an important theoretical tool for a better understanding of UV-IR connections in QFT and, more practically, an essential in-road when modelling and searching for signs of new high-energy physics.

In this work we derive new positivity bounds for massive particles with spin, beyond the forward limit, that are strictly stronger than previous bounds in many cases. The main new ingredient that goes into deriving our bounds is a stronger unitarity condition, or rather a pair of unitarity conditions that constrain both the $s$-channel and $u$-channel branch cuts in terms of the exchanged angular momentum, taking careful account of the non-trivial crossing relations for massive spinning particles.
As a first application of our results, we obtain new constraints on a variety of operators, starting at dimension-6.

The main results of this work are briefly summarised in Section~\ref{sec:results} below, followed by a list of our notation and conventions in Section~\ref{sec:notation}. 
In Section~\ref{sec:unitarity} we derive a new unitarity bound on the UV amplitude using the $s$-channel partial wave expansion, and then in Section~\ref{sec:crossing} we use crossing to derive the analogous bound in the $u$-channel region. 
In Section~\ref{sec:positivity}, we then invoke causality (analyticity) and locality (boundedness) to construct a dispersion relation which relates these UV unitarity constraints to positivity bounds in the IR. Section~\ref{sec:examples} provides various simple examples of low-energy EFTs with massive spinning fields that can be constrained with these new positivity bounds, and we conclude in Section~\ref{sec:discussion} by highlighting future applications and the potential to further develop these positivity bounds.

\subsection{Summary of Main Results}
\label{sec:results}

Our main result is the improved positivity bound \eqref{eqn:intro_bound}, which follows from new, stronger $t$-derivative unitarity conditions for spinning particles in the helicity basis. 
These conditions exploit angular momentum conservation in the UV, which leads to a series of \emph{selection rules} in the partial wave expansion. 
Furthermore, previous beyond-the-forward-limit positivity bounds in the literature have specialised to massless spinning particles or have restricted to transverse spin projections, and here we are able to establish bounds directly in the helicity basis for the first time by carefully considering the crossing relation for massive spinning particles. 

Specifically, we derive the following unitarity condition for the $s$-channel, 
\begin{align} 
 (s-4m^2)  \partial_t \text{Abs}_s \, \mathcal{A}(s, t) \big|_{t=0} &\geq |h_s | \; \text{Abs}_s \, \mathcal{A}(s, t) \big|_{t=0} \; ,
 \label{eqn:unit_dtAs_intro} 
\end{align}
where $h_s=h_1 - h_2$ is the total $s$-channel helicity and the absorptive part $\text{Abs}_s$ of a scattering amplitude $\mathcal{A}$ is defined in \eqref{eqn:Disc_s}. 
Previous unitarity conditions in the literature are equivalent to just the left-hand side of the inequality \eqref{eqn:unit_dtAs_intro} being positive. 
Here \eqref{eqn:unit_dtAs_intro} is more constraining since it is further bounded from below for particles with non-zero helicity, and reflects the fact that only modes with $J \geq |h_s|$ can contribute to the partial wave expansion.  

Positivity only holds if both the integrals across the $s$- and $u$-channel discontinuities are positive. In the massless case, positivity in the $u$-channel follows trivially from the $s$-channel case, since crossing symmetry is trivial. But in the massive case this is not as straightforward. Here we show that the following helicity averaged amplitude, 
\begin{align}
	\mathcal{A}_{h_u} (s,t)  :=  \sum_{ \substack{ h_1 , \,  h_2 \\ h_1 + h_2 = h_u }} \mathcal{A}_{h_1 h_2 h_1 h_2} (s,t)
\end{align}
has a positive $u$-channel discontinuity that is also bounded from below, 
\begin{align}
 (u- 4m^2) \partial_t \, \text{Abs}_u \, \mathcal{A}_{h_u} (4m^2 -u - t , t ) |_{t=0} 
 &\geq
|h_u|  \text{Abs}_u \, \mathcal{A}_{h_u} (4m^2 -u - t , t ) |_{t=0} \; ,
\label{eqn:unit_dtAu_intro}
\end{align}
which follows from the selection rule $J \geq |h_u|$ in the $u$-channel partial wave expansion. 

These results allow us to derive new positivity bounds. To account for unphysical kinematic singularities, the bounds are expressed in terms of a regulated amplitude,
\begin{equation}
	\mathcal{\hat{A}}_{h_u} (s,t) := \frac{ \left(  s ( s - 4m^2  )  \right)^{S_1 + S_2}  }{ \left( - s u  \right)^{|  h_s |_{\rm min} } } \mathcal{A}_{h_u} (s,t) \, ,
	\label{eqn:intro_Ahat}
\end{equation} 
where $S_{1,2}$ are the spins of particles $1,2$. In the low-energy EFT, we may calculate $\partial_s^{2N} \mathcal{\hat{A}}_{h_u}(s_0, t; \mu)$ at $s_0=4m^2$, where all poles and branch cuts are subtracted up to a scale $\mu > 4m^2$ within the EFT. 
This is then related via a dispersion relation to the UV values of the absorptive $\text{Abs}_s$ and $\text{Abs}_u$ parts appearing in \eqref{eqn:unit_dtAs_intro} and \eqref{eqn:unit_dtAu_intro}. 
To obtain a positivity bound, the number $N$ should be chosen depending on the assumption on the UV energy growth, such that the contour integral at infinity vanishes, e.g. locality guarantees this for $N \geq 1 + S_1 + S_2 - |h_s|_\text{min}$ where $|h_s|_\text{min}$ is the minimum value of $|h_1 - h_2|$ in the sum \eqref{eqn:intro_Ahat}. We obtain the following new positivity bounds, 
\begin{align}
	\left.\partial_t \partial_s^{2N} \mathcal{\hat{A}}_{h_u}(4m^2, t;\mu)\right|_{t=0} > 
	\left\{ \begin{array}{cc} 
		-\frac{\alpha}{\mu}\partial_s^{2N} \mathcal{\hat{A}}_{h_u}(4m^2, 0;\mu) & \, , \quad \alpha > 0 \\
		0 & \, , \quad \alpha \leq 0 
	\end{array} \right. \, ,
	\label{eqn:intro_bound}
\end{align}
where we have defined 
\begin{equation}
	\alpha := 2(N-S_1-S_2 + |h_s|_\text{min} )-|h_u|-|h_s|_\text{min} \, .
\end{equation}
When $\alpha > 0$, \eqref{eqn:intro_bound} reproduces the known scalar positivity bounds when $h_1=h_2=0$ and is qualitatively stronger than existing bounds in the literature for non-zero helicities. The bound when $\alpha \leq 0$ is qualitatively different than existing bounds because it bounds the first $t$-derivative independently of the zeroth $t$-derivative. 

As with other positivity bounds in the literature, the bound \eqref{eqn:intro_bound} only applies if the high-energy growth of the amplitude in the UV theory is assumed to be sufficiently bounded.
Typically this is ensured by the Froissart bound, $\lim_{|s|\to \infty} | \mathcal{A} (s,t) | < s^2$, which follows from unitarity and causality in any local quantum theory with a mass gap. 
In this case, 
the positivity bound \eqref{eqn:intro_bound} applies for all $2 (N - S_1 -S_2 + |h_s|_{\rm min} ) \geq 2$. 
This corresponds to constraining interactions of mass dimension-8 and higher. Constraining lower dimension operators requires stronger assumptions about the UV growth. For instance if $\lim_{|s| \to \infty} | \mathcal{A} (s,t) | < s^0$ in the UV then the bound with $2 (N - S_1 -S_2 + |h_s|_{\rm min} ) = 0$ applies and can be used to constrain dimension-6 operators whose amplitudes contain a piece that grows as $s^0 t^1$. 

One interesting distinction between~\eqref{eqn:intro_bound} and previous positivity bounds is that when $\alpha \leq 0$ (i.e. for sufficiently large $|h_u| + |h_s|_{\rm min}$) the first $t$-derivative is constrained independently of the forward-limit amplitude without any $t$-derivatives.
As a result, this $\alpha \leq 0$ bound applies given weaker assumptions about the UV growth.
In particular, dimension-6 operators can now be constrained with the milder assumption that $\lim_{|s| \to \infty} \partial_t | \mathcal{A} (s,t) | < s^0$.

\subsection{Notation and Conventions} 
\label{sec:notation}

Here we set out the notations and conventions that we use in the rest of the paper.
We work in $3+1$ spacetime dimensions with metric signature $(- +++)$.  

\paragraph{Two-particle states:}
A single-particle state $| \bfp \, Q \rangle$ is described by the particle's spatial momentum $\bfp$ and its quantum numbers $Q$, which include the spin, $S$, and helicity, $h$, as well as other identifiers like species, flavour, charge, etc. 
The most important of these in this work will be the particle helicity, and so we adopt the shorthand $| \bfp \, h \rangle$, where only the helicity is written explicitly\footnote{
Other quantum numbers are implicit, e.g. the state $|\bfp_1 \, h_1 \rangle$ of particle 1 implicitly has spin $S_1$. 
}.
A two-particle state is then written as the product $|\bfp_1 \, h_1 \rangle | \bfp_2 \, h_2 \rangle$. 

We adopt the usual relativistic normalisation of the one-particle state,
\begin{align}
\langle \bfp' h' | \bfp h \rangle = 2 \omega_p \delta_{h h'} (2\pi)^3 \delta^3 ( \bfp - \bfp' ) 
\end{align}
where $\omega_p = \sqrt{ | \bfp|^2 + m^2}$ is the energy. This ensures that the corresponding $2$-particle state $| \bfp_1 h_1 \rangle | \bfp_2 h_2 \rangle$ has normalisation, 
\begin{align}
\int d^6 \Pi_{p_1 p_2}  \langle \bfp_2' h_2' | \langle \bfp_1' h_1' | \bfp_1 h_1 \rangle | \bfp_2 h_2 \rangle = \delta_{h_1 h_1'} \delta_{h_2 h_2'} 
\label{eqn:p1p2_norm}
\end{align}
with respect to the Lorentz-invariant two-particle phase space element,
\begin{align}
d^6 \Pi_{p_1 p_2 } = \frac{1}{g_s}  d^3 \Pi_{p_1} d^3 \Pi_{p_2}   \; , \;\;\;\;  d^3 \Pi_{p} =  \frac{d^4 p}{(2 \pi )^4} \,  2 \pi \delta \left( p^2 - m^2 \right) \Theta ( p_0 ) \, ,
\label{eqn:dPi}
\end{align}
which corresponds to integrating over all on-shell, future-pointing momenta for each particle, and dividing by a degeneracy factor $g_s$ to account for whether the two ingoing particles are distinguishable $(g_s=1)$ or indistinguishable ($g_s=2)$ in the $s$-channel. This $g_s$ factor is required since when indistinguishable the two-particle phase space is a factor of 2 smaller (because $| \bfp_1 \bfp_2 \rangle$ and $|\bfp_2 \bfp_1 \rangle$ are identified).

\paragraph{$\bm{2 \to 2}$ amplitudes:}
Since we assume Lorentz invariance throughout, the $2 \to 2$ scattering amplitude can be expressed in terms of the usual Mandelstam variables\footnote{
$p_a$ is the usual 4-momentum $(\pm \omega_{p_a} , \bfp_a)$, where the sign is $+$ for ingoing and $-$ for outgoing.
},
\begin{align}
 s= - ( p_1 + p_2)^2 \;\;\;\; , \;\;\;\; t = - ( p_1 + p_3 )^2 \;\;\;\; , \;\;\;\; u = - (p_1 + p_4 )^2 \; ,
 \label{eqn:stu}
\end{align}
which are related by $s+t+u = \sum_a m_a^2$.  
We will focus on elastic scattering processes in which $m_1 = m_3$, $S_1 = S_3$ and $m_2 = m_4$, $S_2 = S_4$. In the main text, we make the further simplifying assumption that $m_1 = m_2$, so that all four particles have identical mass, $m$---this is largely for cosmetic reasons, and we give in Appendix~\ref{app:unequal} the analogous derivation for the $m_1 \neq m_2$ case. 

We will refer to the process  $12 \to 34$ as ``the $s$-channel'' and the process $1 \bar{4} \to 3 \bar{2}$ as ``the $u$-channel''. The bar denotes the corresponding anti-particle quantum numbers, for instance $\bar{h} = - h$ for the helicities. 
Physical $s$-channel kinematics (i.e. real ingoing $p_1, p_2$ and outgoing $p_3, p_4$) corresponds to the region $s - 4m^2 > -t > 0$, and physical $u$-channel kinematics (real ingoing $p_1, p_4$ and outgoing $p_3, p_2$) corresponds to the region $s \leq 0 \leq -t$.
When discussing the elastic helicity configuration $h_1 = h_3$ and $h_2 = h_4$, we will often refer to the $s$- and $u$-channel total helicities,
\begin{align}
h_s := h_1 - h_2 \;\; , \;\;\;\; h_u := h_1 - \bar{h}_4 =  h_1 + h_2 \; . 
\end{align}

We use $\mathcal{A}_{h_1 h_2 \to h_3 h_4} ( \bfp_1 , \bfp_2 , \bfp_3 , \bfp_4 )$ to denote the amplitude that $|\bfp_1 \, h_1 \rangle | \bfp_2 \, h_2 \rangle$ will transition to $| \bfp_3 \, h_3 \rangle | \bfp_4 \, h_4 \rangle$ (i.e. the physical on-shell $S$-matrix element), and we use $\mathcal{A}_{h_1 h_2 h_3 h_4} (s,t)$ to denote the analytic continuation of this transition amplitude beyond the physical $s$-channel region to complex values of $s$ and $t$.
Similarly, we use $\mathcal{A}_{h_1 \bar{h}_4 \to h_3 \bar{h}_2} ( \bfp_1 , \bfp_4 , \bfp_3 , \bfp_3 )$ to denote the amplitude that $|\bfp_1 \, h_1 \rangle | \bfp_4 \, \bar{h}_4 \rangle$ will transition to $| \bfp_3 \, h_3 \rangle | \bfp_2 \, \bar{h}_2 \rangle$, and use $\mathcal{A}_{h_1 \bar{h}_4 h_3 \bar{h}_2} (u,t)$ to denote its analytic continuation beyond the physical $u$-channel region. The bars on the helicity indices make it clear whether an amplitude refers to the $s$- or the $u$-channel process, but where it may be ambiguous we include a superscript to denote the channel, e.g. $\mathcal{A}^{\psi_1 \psi_2 \to \psi_3 \psi_4}_{h_1 h_2 \to h_3 h_4} (s,t)$ for the process in which fields $\psi_1 \psi_2$ are ingoing and $\psi_3 \psi_4$ are outgoing.

\paragraph{$\bm{s}$-channel kinematics:}
Each value of $s$ and $t$ corresponds to a family of possible particle momenta $p_a^\mu$ (which differ from one another by an overall Lorentz transformation). 
In a Lorentz-invariant theory, each of these possible sets of momenta are physically equivalent, however it will nonetheless be useful to choose a canonical frame in which to work. 
In the $s$-channel region, $s \geq 4m^2 - t \geq 4m^2$, we use the centre-of-mass momenta\footnote{
Note that although we have written the 4-momenta of all particles as incoming, so that $\sum_a p_a = 0$, the states are labelled by the physical momenta of the particles, i.e. the outgoing $\langle \bfp_3 \, h_3 |$ particle has 3-momentum $\bfp_3 = +\tfrac{1}{2} \left( k_s \sin \theta_s , 0 , k_s \cos \theta_s \right)$. 
},
\begin{align}
 p_1^s = \tfrac{1}{2} \left( \begin{array}{c}
 \sqrt{s} \\
 0 \\
 0 \\
 k_s
 \end{array} \right) \;\; , \;\; 
  p_2^s = \tfrac{1}{2}  \left( \begin{array}{c}
  \sqrt{s} \\
 0 \\
 0 \\
- k_s
 \end{array} \right) \;\; , \;\; 
  p_3^s = \tfrac{1}{2} \left( \begin{array}{c}
  - \sqrt{s} \\
 - k_s \sin \theta_s \\
 0 \\
 - k_s \cos  \theta_s
 \end{array} \right)  \;\; , \;\; 
  p_4^s =   \tfrac{1}{2} \left( \begin{array}{c}
-  \sqrt{s} \\ 
 k_s \sin \theta_s \\
 0 \\
  k_s \cos \theta_s
 \end{array} \right)
  \label{eqn:kinematics}
\end{align}
where $k_s = \sqrt{s-4m^2}$, and the scattering angle $\theta_s$ is given by,
 \begin{align}
 \cos \frac{\theta_s}{2} = \frac{ \sqrt{-u} }{ \sqrt{s-4m^2} } \;\;\;\; , \;\;\;\;  \sin \frac{\theta_s}{2} = \frac{ \sqrt{-t} }{ \sqrt{s-4m^2} } \; . 
 \label{eqn:ths}
\end{align}
This has the advantage that $\mathcal{A} ( k_s , \theta_s)$ describes physical scattering for any real $k_s \geq 0$ and any real $\theta_s$, and in particular allows for a partial wave expansion in which $\theta_s$ is transformed into an angular momentum, $\mathcal{A}_{\ell} (s)$.

\paragraph{$\bm{u}$-channel kinematics:}
Similarly, in the $u$-channel region, we use the $u$-channel centre-of-mass momenta,
\begin{align}
 p_1^u =  \tfrac{1}{2} \left( \begin{array}{c}
 \sqrt{u} \\
 0 \\
 0 \\
 k_u
 \end{array} \right) \;\; , \;\; 
  p_4^u = \tfrac{1}{2}  \left( \begin{array}{c}
  \sqrt{u} \\
 0 \\
 0 \\
- k_u
 \end{array} \right) \;\; , \;\; 
  p_3^u = \tfrac{1}{2} \left( \begin{array}{c}
  - \sqrt{u}  \\
 - k_u \sin \theta_u \\
 0 \\
 - k_u \cos  \theta_u
 \end{array} \right)  \;\; , \;\; 
  p_2^u =  \tfrac{1}{2} \left( \begin{array}{c}
 -\sqrt{u}  \\ 
 k_u \sin \theta_u \\
 0 \\
  k_u \cos \theta_u
 \end{array} \right)
 \label{eqn:kinematics_u}
\end{align}
where $k_u = \sqrt{u-4m^2}$ and the scattering angle is,
\begin{align}
 \cos \frac{\theta_u}{2} = \frac{ \sqrt{-s} }{ \sqrt{u-4m^2} } \;\;\;\; , \;\;\;\;  \sin \frac{\theta_u}{2} = \frac{ \sqrt{-t} }{ \sqrt{u-4m^2} } \; , 
 \label{eqn:thu}
\end{align}
for which there is an analogous partial wave expansion. 

Finally, the $s$- and $u$-channel partial wave expansions are related to each other by crossing symmetry, which is most easily described using a further angular variable,
\begin{align}
 \cos \, \chi_u &=  \frac{ \sqrt{ - s u } }{ \sqrt{ -  (s-4m^2)  (u-4m^2) } }   = \sqrt{ \frac{u (u-4m^2)}{s (s-4m^2)} }  \; \frac{1 + \cos \theta_u}{2}    \;\;, \nonumber \\
 \sin \chi_u &= \frac{- 2 m \sqrt{-t} }{ \sqrt{ - (s-4m^2) (u-4m^2) }  }   = \sqrt{ \frac{u (u-4m^2)}{s (s-4m^2)} } \; \frac{m}{\sqrt{u}}  \sin \theta_u
 \label{eqn:chiu}
\end{align} 
which corresponds to the angle through which $p_2$ rotates upon boosting from the Lorentz frame \eqref{eqn:kinematics} (in which $p_1^\mu + p_2^\mu = \sqrt{s} \, \delta^\mu_0$) to \eqref{eqn:kinematics_u} (in which $p_1^\mu + p_4^\mu = \sqrt{u} \, \delta^\mu_0$).

\section{Unitarity Bounds with Angular Momentum}
\label{sec:unitarity}

In this section, we derive the $s$-channel unitarity bound \eqref{eqn:unit_dtAs_intro}, which bounds the $t$ derivative of the scattering amplitude in terms of the forward limit amplitude in the region $s-4m^2 \geq -t \geq 0$.

\paragraph{Unitarity and the Optical Theorem:}
Unitarity of the $\hat{S}$-matrix, $\hat{S}^\dagger \hat{S} = \mathbbm{1}$, is required by the conservation of probability: it ensures that the norm of the wavefunction does not change with time. 
This is a cornerstone of quantum field theory. 
Separating $\hat{S}$ into free and interacting parts, $\hat{S} = \mathbbm{1} + i \hat{T}$, 
unitarity requires that,
\begin{align}
\hat{T} - \hat{T}^\dagger = i \hat{T}^\dagger \hat{T} \; ,
\label{eqn:unit_T}
\end{align}
which loosely speaking constrains the ``imaginary part'' of any scattering amplitude in terms of its absolute value. 
More explicitly, the 2 $\to$ 2 scattering amplitude $\mathcal{A}$ is related to an $\hat{S}$-matrix element via,
\begin{align}
 \langle \bfp_4 \, h_4 | \langle \bfp_3 \, h_3 | \hat{T} | \bfp_1 \, h_1 \rangle | \bfp_2 \, h_2 \rangle  = (2\pi)^4 \delta^4 (\bfp_1 + \bfp_2 - \bfp_3 - \bfp_4 ) \, \mathcal{A}_{h_1 h_2 \to h_3 h_4} 
 \; .  
 \label{eqn:Adef}
\end{align}
Although \eqref{eqn:Adef} only defines $\mathcal{A}_{h_1 h_2 \to h_3 h_4}$ for real values of $s$ and $t$ in the region $s-4m^2 \geq -t \geq 0$ (which corresponds to all four $\bfp_a^s$ being real in \eqref{eqn:kinematics}), we can define the analytic continuation $\mathcal{A}_{h_1 h_2 h_3 h_4} (s,t)$ to all complex values of $s$.
The left-hand-side of \eqref{eqn:unit_T} then corresponds to the ``absorptive'' part of this amplitude\footnote{
Note $\text{Abs}_s \, f(s) = \tfrac{1}{2i} \text{Disc} \, f(s)$ is the usual discontinuity of a complex function across a branch cut. 
}, 
\begin{align}
\text{Abs}_s \; \mathcal{A}_{h_1 h_2 h_3 h_4} ( s  , t )  &:=  \tfrac{1}{2i} \lim_{\epsilon \to 0} \left(  \mathcal{A}_{h_1 h_2  h_3 h_4} (s + i \epsilon, t) - \mathcal{A}_{h_1 h_2 h_3 h_4} (s - i \epsilon ,t )  \right)   \nonumber \\
&:=  \tfrac{1}{2 i } \left( \mathcal{A}_{h_1 h_2 \to h_3 h_4} - \mathcal{A}_{h_3 h_4 \to h_1 h_2}^*   \right) \;\; \text{when } s - 4m^2 \geq -t \geq 0 \; .
\label{eqn:Disc_s}
\end{align}
which reduces to $\text{Im} \, \mathcal{A}$ for processes which are invariant under time-reversal\footnote{
Note that $\mathcal{A}_{h_1 h_2 h_3 h_4} (s - i \epsilon ,t ) = \mathcal{A}_{h_3 h_4 h_1 h_2}^* (s + i \epsilon , t) $ due to Hermitian analyticity of the $\hat{S}$ ($\hat{T}$) matrix, and we define the analytic continuation $\mathcal{A}_{h_1 h_2 h_3 h_4} (s  ,t )$ in the complex $s$-plane such that it coincides with the matrix element $\mathcal{A}_{h_1 h_2 \to h_3 h_4}$ on the real $s$-axis approached from above. 
}. 
Then if we insert a complete set of $N$ particles states on the right-hand-side of \eqref{eqn:unit_T}, unitarity becomes the celebrated optical theorem,
\begin{align}
2\,  \text{Abs}_s \, \mathcal{A}_{h_1 h_2 h_3 h_4} (s, t) = \sum_N  \mathcal{A}_{h_1 h_2 \to N} \mathcal{A}_{h_3 h_4 \to N}^*   \;\;\;\; \text{when} \;\;  s - 4m^2 \geq -t \geq 0 \; .
\label{eqn:optical}
\end{align}
In particular, for \emph{elastic processes} ($h_3 = h_1$ and $h_4 = h_2$) in the \emph{forward limit} ($\bfp_3 = \bfp_1$ and $\bfp_4 = \bfp_2$) then the right-hand-side of \eqref{eqn:optical} becomes $\sum_N | \mathcal{A}_{h_1 h_2 \to N} |^2$ and is always strictly positive in an interacting theory.
Although not manifest from \eqref{eqn:optical}, this positivity also extends beyond the forward limit \cite{Nicolis:2009qm}, allowing positivity bounds to be placed on all $t$ derivatives of the amplitude \cite{deRham:2017avq, deRham:2017zjm} (see also \cite{Vecchi:2007na, Pennington:1994kc, Manohar:2008tc} for earlier applications). This is made possible using the partial wave expansion, as we will now show (see e.g. \cite{Richman:1984gh} for a more detailed review).
 
\paragraph{Partial wave expansion:}
The main idea behind the partial wave expansion is to expand the incoming/outgoing two-particle states in terms of the states $ | P_\mu \; J \; M \rangle | Q \rangle$, where $P_\mu$ is the eigenvalue of spacetime translations and $(J,M)$ are the total and magnetic angular momentum eigenvalues, while
$Q$ are all of the \emph{internal} quantum numbers which commute with spacetime translations and rotations. 
The virtue of these states is that the conservation rules associated with the spacetime isometries are made manifest, since the interactions commute with spacetime translations and Lorentz transformations\footnote{
Note that $\hat{T} (P^2, J , M)$ retains a hat because \eqref{eqn:T_partial} represents only a partial trace over the external quantum numbers, leaving an operator which acts on the subspace of internal quantum numbers (e.g. for the 2-particle case \eqref{eqn:state_pw0}, $\hat{T} (P^2, J, M)$ acts on the $|h_1 h_2 \rangle$ part of the states).
},
\begin{align}
 \langle P_{\mu}' \; J' \; M' | \hat{T} | P_\mu \; J \; M \rangle = (2\pi)^4 \delta^4 \left( P_\mu - P_{\mu}' \right) \delta_{J J'} \delta_{M M'} \, \hat{T} ( P^2 , J , M )
 \label{eqn:T_partial}
\end{align}
  
Explicitly, the incoming two-particle state $|\bfp_1 \, h_1 \rangle | \bfp_2 \, h_2 \rangle$ (where we are only writing the helicity labels explicitly), can be expanded in the $s$-channel region as, 
\begin{align}
 | \bfp_1 \, h_1 \rangle | \bfp_2 \, h_2 \rangle = \sum_{J_s = 0}^{\infty} \sum_{M_s = -J_s}^{+J_s} \; c_{J_s M_s}  \,  | p_s \; J_s \; M_s \rangle | h_1 \, h_2 \rangle 
 \label{eqn:state_pw0}
\end{align} 
where $p_s = p_1 + p_2$ is the total incoming 4-momentum in the $s$-channel, and $J_s$, $M_s$ are the incoming angular momentum quantum numbers. 
In particular, $M_s$ is the eigenvalue of $\hat{J}_z =  \hat{L}_z + \hat{S}_z^{(1)} + \hat{S}_z^{(2)}$, and represents the sum of the orbital angular momentum associated with the relative motion of particles 1 and 2 and their intrinsic spins. 
These states are normalised so that,
\begin{align}
\langle p_s' \, J_s' \, M_s' | p_s \, J_s \, M_s \rangle = V_s ( 2 \pi )^4 \delta^4 ( p_s - p_s' ) \delta_{J_s J_s'} \delta_{M_s M_s'} \;\; \text{and} \;\;    \langle h_1' h_2' | h_1 h_2 \rangle = \delta_{h_1 h_1'} \delta_{h_2 h_2'} 
\label{eqn:PJM_norm}
\end{align}
where $V_s$ is the relative phase space volume between $| \bfp_1 \; \bfp_2 \rangle$ and $| p_s \; J_s \; M_s \rangle$, given by $V_s = 8 \pi g_s (2 J_s +1 ) \sqrt{s} / k_s$ for the centre-of-mass kinematics \eqref{eqn:kinematics}\footnote{
In a general Lorentz frame, this relative phase space factor is given explicitly by,
\begin{align}
 V_s :=  (2J_s+1) \frac{ d^4 p_s  }{ d^6 \Pi_{p_1 p_2} } \frac{d^2 \Omega_{p_1} }{4 \pi }  \; ,
\end{align}
where $d^2 \Omega_{p_1}$ are the two spherical angles of $p_1$ and $d^6 \Pi_{p_1 p_2}$ is given in \eqref{eqn:dPi}.
}. 

To find the coefficients $c_{J_s M_s}$ in \eqref{eqn:state_pw0}, we first use the fact that $\hat{L}_z$ vanishes when $\bfp_1$ and $\bfp_2$ collide along the $z$-axis, in which case $M_s = h_1 - h_2$ is simply equal to the total ingoing helicity\footnote{
Note that $h_2$ contributes negatively to the total helicity since particle 2 is travelling in the opposite direction to particle 1, i.e. along the negative $z$-axis. 
}, a combination we will define as $h_s^{\rm in} := h_1 - h_2$. 
Then comparing the normalisation \eqref{eqn:PJM_norm} with \eqref{eqn:p1p2_norm} fixes each $c_{J_s h_s^{\rm in}} = 1$ up to an unimportant phase. 
Altogether, the incoming 2-particle state can be expanded as,
\begin{align}
 | \bfp_1^s \, h_1 \rangle | \bfp_2^s \, h_2 \rangle = \sum_{J_s = |h^{\rm in}_s| }^{\infty} \;   | p_s \; J_s  \; h_s^{\rm in} \rangle | h_1 \, h_2 \rangle \; . 
 \label{eqn:state_pw}
\end{align} 
Note that the sum begins at $J_s = | h_s^{\rm in} |$ since $J_s < M_s$ is forbidden. 

Defining the angular momentum $\hat{J}_{y}$ as the generator of rotations within the scattering plane (i.e. recall that with the conventions \eqref{eqn:kinematics} all particle momenta lie within the $xz$-plane), then we can similarly write for the outgoing state,
\begin{align}
 | \bfp_3^s \, h_3 \rangle | \bfp_4^s \, h_4 \rangle = \sum_{J_s = |h_s^{\rm out}| }^{\infty} \; e^{i \hat{J}_{y} \theta_s }  | p_s \; J_s  \; h_s^{\rm out} \rangle | h_3 \, h_4 \rangle 
 \label{eqn:state_pw2}
\end{align} 
where $h_s^{\rm out} := h_3 - h_4$, which follows from \eqref{eqn:state_pw} since the outgoing momenta ($\bfp_3^s$, $\bfp_4^s$) are related to the ingoing ($\bfp_1^s$, $\bfp_2^s$) by a rotation of the scattering plane by an angle $\theta_s$. 

Finally, in the $s$-channel region where $\mathcal{A}_{h_1 h_2 h_3 h_4} (s,t) = \mathcal{A}_{h_1 h_2 \to h_3 h_4} \left( \bfp_1^s, \bfp_2^s, \bfp_3^s, \bfp_4^s  \right)$, we can substitute \eqref{eqn:state_pw} and \eqref{eqn:state_pw2} into the matrix element definition of the amplitude \eqref{eqn:Adef} to write,
\begin{align}
 \mathcal{A}_{h_1 h_2 h_3 h_4} (s, t)
 &=
 \sum_{J_s }   \langle J_s h_s^{\rm out} | e^{-i \hat{J}_{y} \theta_s}  |  J_s h_s^{\rm in} \rangle \;  \langle h_3 h_4 | \hat{T} (s, J_s , h_s^{\rm in} ) | h_1 h_2 \rangle   \;  , 
 \label{eqn:Apw}
\end{align}
where the sum over $J_s$ starts at $\text{max} \left( |h_s^{\rm in}| , | h_s^{\rm out}| \right)$. \eqref{eqn:Apw} is the partial wave expansion for massive spinning particles\footnote{
Note that $ \langle J_s h_s^{\rm out} | e^{-i \hat{J}_{y} \theta_s}  |  J_s h_s^{\rm in} \rangle = d^{J_s}_{h_s^{\rm out} h_s^{\rm in}} ( \theta_s)$ is Wigner's $d$ matrix, which reduces to the usual Legendre polynomials for spin-less particles, $d^J_{00} ( \theta) = P_J ( \cos \theta)$.
}, which we can now use to establish unitarity bounds on $\text{Abs}_s \, \mathcal{A}$ and its $t$ derivatives.

\paragraph{Forward limit:}
In the forward limit, $t=0$ (which implies $\theta_s = 0$), the $\hat{J}_y$ matrix element in \eqref{eqn:Apw} becomes a simple $\delta$-function imposing conservation of angular momentum about the $z$-axis, $\delta_{h_s^{\rm in} h_s^{\rm out} }$. 
Then taking the $\text{Abs}_s$ part, and using unitarity \eqref{eqn:unit_T} to replace $\hat{T}$ with $\hat{T}^\dagger \hat{T}$, we can write,
\begin{align}
& 2 \,  \text{Abs}_s \, \mathcal{A}_{h_1 h_2 h_1 h_2} (s , 0)  =  \sum_{J_s = | h_s |}^{\infty}  \left|  \langle T_s | h_1 h_2  \rangle \right|^2   \; . 
\label{eqn:DiscAs_fwd}
\end{align}
where we have introduced $|T_s \rangle$ as a convenient shorthand for the partial trace,
\begin{align}
 \langle p_s \; J_s \; h_s | \hat{T}^\dagger \hat{T} | p_s \; J_s \; h_s \rangle = | T_s \rangle \langle T_s | \; . 
\end{align}
To compute $\langle T_s | h_1 h_2  \rangle$ explicitly requires complete knowledge of $\hat{T}$ in the interacting UV theory. From an EFT perspective this is often not possible, however crucially this quantity is always sign definite, and so, 
\begin{align}
  \text{Abs}_s \, \mathcal{A}_{h_1 h_2 h_1 h_2} (s , 0 )  > 0  \; ,
  \label{eqn:unit_As}
\end{align}
for all $s-4m^2 \geq  0$ is a robust prediction of unitarity for \emph{any} possible interactions.

\paragraph{First $\bm{t}$ derivative:}
Since $(s-4m^2)\partial_t = 2 \partial / \partial  \cos \theta_s$ ($\sim - 2 \partial_{\theta_s}^2$ at $\theta_s=0$), we find that for elastic processes in the $s$-channel region, 
\begin{align}
 (s-4m^2)  \partial_t \, \text{Abs}_s \, \mathcal{A}_{h_1 h_2 h_1 h_2} (s, t) \big|_{t=0} =   \sum_{J_s =  |h_s|  }^{\infty} \langle J_s h_s |  \hat{J}_y^2  | J_s h_s  \rangle \;  | \langle T_s | h_1 h_2 \rangle |^2 \; . 
 \label{eqn:dtDiscA}
\end{align}
Since $\hat{J}_{y}$ is Hermitian, it is clear that the right-hand-side of \eqref{eqn:dtDiscA} is positive definite. 
Indeed, this is how positivity bounds on $t$ derivatives were established in \cite{deRham:2017avq, deRham:2017zjm}. 
The step forward which we wish to take in this work is the observation that this angular momentum matrix element is not only positive, but is also \emph{bounded from below}, 
\begin{align}
2 \langle J_s\,  h_s | \hat{J}_{y}^2 | J_s \, h_s \rangle  &=   J_s (J_s + 1 ) - h_s^2  \geq |h_s| \; .
\label{eqn:Jy2_identity}
\end{align}
Physically, this is simply a consequence of angular momentum conservation: since the ingoing angular momentum along the $z$-axis is $|h_s|$, the total angular momentum $J_s$ of any partial wave contribution must be $\geq |h_s|$.   
Comparing \eqref{eqn:dtDiscA} with \eqref{eqn:DiscAs_fwd}, we find that the selection rule \eqref{eqn:Jy2_identity} leads to the following bound, \vskip5pt
\begin{mdframed}[style=boxed]  \vskip-20pt
\begin{align} 
 (s-4m^2)  \partial_t \, \text{Abs}_s \, \mathcal{A}_{h_1 h_2 h_1 h_2} (s, t) \big|_{t=0} \geq |h_s | \; \text{Abs}_s \, \mathcal{A}_{h_1 h_2 h_1 h_2} (s, 0)  \; ,
 \label{eqn:unit_dtAs}
\end{align} 
\end{mdframed}
in the physical $s$-channel region, $s-4m^2 \geq 0$. 
In fact, since higher order $t$-derivatives can be related to the matrix elements $\langle J_s \, h_s | \hat{J}_y^{2n} | J_s \, h_s \rangle$, which obey analogous selection rules to \eqref{eqn:Jy2_identity}, the bound \eqref{eqn:unit_dtAs} is the first in an infinite tower of such inequalities, which we construct explicitly in Appendix~\ref{app:higher}. 

~

The unitarity bound \eqref{eqn:unit_dtAs} is the first ingredient of the positivity bounds that we derive in this paper: it establishes a bound on the discontinuity of the right-hand ($s$-channel) branch cut of $\mathcal{A}_{h_1 h_2 h_1 h_2}$ which is strictly stronger than the basic positivity condition $\partial_t \text{Abs}_s \, \mathcal{A}_{h_1 h_2 h_1 h_2} > 0$ when $|h_s| \neq 0$. 
The second ingredient we require is the $u$-channel analogue of \eqref{eqn:unit_dtAs}, in order to place a bound on the discontinuity of the left-hand branch cut. This is achieved by applying a crossing transformation to $\mathcal{A}_{h_1 h_2 h_1 h_2} (s,t)$---unlike for scalar particles, this amplitude carries little group indices and so transforms non-trivially under crossing, which we will now describe in detail.

\section{Crossing Relation with Spin}
\label{sec:crossing}

In this section, we derive the $u$-channel unitarity bound \eqref{eqn:unit_dtAu_intro}, which bounds the $t$ derivative of the scattering amplitude in terms of the forward limit amplitude in the region $s \leq 0 \leq -t$ (i.e. in the $u$-channel region, $u-4m^2 \geq -t \geq 0$).

\paragraph{$\bm{u}$-channel unitarity:}
The scattering amplitude for the $u$-channel process, $\mathcal{A}_{h_1 \bar{h}_4 \to h_3 \bar{h}_2}$, is defined by the matrix element, 
\begin{align}
 \langle \bfp_2 \, \bar{h}_2 | \langle \bfp_3 \, h_3 | \hat{T} | \bfp_1 \, h_1 \rangle | \bfp_4 \, \bar{h}_4 \rangle  = (2\pi)^4 \delta^4 (\bfp_1 + \bfp_4 - \bfp_3 - \bfp_2 ) \mathcal{A}_{h_1 \bar{h}_4 \to h_3 \bar{h}_2}  \; , 
 \label{eqn:Audef}
\end{align}
where the overbar denotes that these quantum numbers belong to the corresponding antiparticle (in particular, the helicity changes sign\footnote{
Since we only write the helicity quantum number explicitly, we are also using the bars over the helicities to implicitly denote the scattering process. In particular, $\mathcal{A}_{h_1 \bar{h}_4 h_3 \bar{h}_2} (= \mathcal{A}^{\psi_1 \bar{\psi}_4 \to \psi_3 \bar{\psi}_2}_{h_1 \bar{h}_4 h_3 \bar{h}_2})$ is generally \emph{not} equal to $\mathcal{A}_{h_1 h_2' h_3 h_4'} (= \mathcal{A}^{\psi_1 \psi_2 \to \psi_3 \psi_4}_{h_1 h_2' h_3 h_4'})$ with $h_2' = -h_4$ and $h_4' = -h_2$ since they correspond to different processes (although they are related to one another, by the crossing relation~\eqref{eqn:helicityCrossing}).
}, $\bar{h}_a = - h_a$). 
This represents a different physical process, in which particle 2 (4) has been exchanged with an outgoing (incoming) antiparticle.
The centre-of-mass energy for this scattering is $u$, and we define $\mathcal{A}_{h_1 \bar{h}_4 h_3 \bar{h}_2} (u,t)$ to be the analytic continuation of $ \mathcal{A}_{h_1 \bar{h}_4 \to h_3 \bar{h}_2}$ to complex values of $u$, exactly analogous to how the $s$-channel amplitude $\mathcal{A}_{h_1 h_2 h_3 h_4} (s,t)$ is defined from \eqref{eqn:Adef}. 
In particular, we can define the absorptive part with respect to $u$,
\begin{align}
\text{Abs}_u \; \mathcal{A}_{h_1 \bar{h}_4 h_3 \bar{h}_2} ( u  , t )  &:=  \tfrac{1}{2i} \lim_{\epsilon \to 0} \left( \mathcal{A}_{h_1 \bar{h}_4 h_3 \bar{h}_2}  (  u + i \epsilon , t ) - \mathcal{A}_{ h_1 \bar{h}_4 h_3 \bar{h}_2 } (  u -  i \epsilon    , t ) \right)   \nonumber \\
&:=  \tfrac{1}{2 i } \left(  \mathcal{A}_{h_1 \bar{h}_4 \to h_3 \bar{h}_2} -  \mathcal{A}_{h_3 \bar{h}_2 \to h_1 \bar{h}_4}^*   \right) \;\; \text{when } u - 4m^2 \geq -t \geq 0 \; ,  
\label{eqn:Disc_u}
\end{align}
so that the unitarity condition \eqref{eqn:unit_T} implies a $u$-channel optical theorem, namely $2 \text{Abs}_u \,\mathcal{A}_{h_1 \bar{h}_4 h_3 \bar{h}_2} (u,t) = \sum_N \mathcal{A}_{h_1 \bar{h}_4 \to N} \mathcal{A}_{h_3 \bar{h}_2 \to N}^*$.
As with the $s$-channel optical theorem, although this immediately implies that $\text{Abs}_u \, \mathcal{A}_{h_1 \bar{h}_2 h_1 \bar{h}_2} (u,0) > 0$ for elastic processes in the forward limit, to extend this positivity beyond forward kinematics requires a partial wave expansion. 
Following the same steps which led to \eqref{eqn:Apw} in the $s$-channel, 
one can show that in the $u$-channel region where $\mathcal{A}_{h_1 \bar{h}_4 h_3 \bar{h}_2} (u,t) =  \mathcal{A}_{h_1 \bar{h}_4 \to h_3 \bar{h}_2} ( \bfp_1^u, \bfp_4^u, \bfp_3^u, \bfp_2^u)$, expanding the 2-particle states in \eqref{eqn:Audef} in terms of $|p_u\, J_u \, M_u \rangle |h_1 \bar{h}_4 \rangle$ leads to, 
\begin{align}
  \mathcal{A}_{h_1 \bar{h}_4 h_3 \bar{h}_2} (u, t) 
 &=
\sum_{J_u }  \langle J_u h_u^{\rm out} | e^{-i \hat{J}_{y} \theta_u}  |  J_u h_u^{\rm in} \rangle 
\;
\langle h_3 \bar{h}_2 | \hat{T}  ( u, J_u , h_u^{\rm in}  ) |  h_1 \bar{h}_4 \rangle 
 \label{eqn:Au_pw}
\end{align}
where $h_u^{\rm in} = h_1 - \bar{h}_4$, $h_u^{\rm out} = h_3 - \bar{h}_2$, and the $J_u$ sum starts at $\text{max} \left( |h_u^{\rm in}| , |h_u^{\rm out}| \right)$. As a result, $\text{Abs}_u \, \mathcal{A}_{h_1 \bar{h}_2 h_1 \bar{h}_2} (u,t)$ satisfies unitarity conditions analogous to \eqref{eqn:unit_As} and \eqref{eqn:unit_dtAs}.

\paragraph{Crossing relation:}
To make use of these $u$-channel unitarity bounds, we must relate the $u$-channel amplitude $\mathcal{A}_{h_1 \bar{h}_4 h_3 \bar{h}_2} (u,t)$ to our original amplitude $\mathcal{A}_{h_1 h_2 h_3 h_4} (s,t)$. 
We do this using \emph{crossing}, which is the property\footnote{
For local quantum theories with a mass gap, crossing has been rigorously proven from unitarity, causality and locality at the level of off-shell correlation functions \cite{Bros:1964iho, Bros:1965kbd}---see \cite{Mizera:2021ujs, Mizera:2021fap} for recent progress towards an entirely on-shell demonstration.
} that the $s$-channel matrix element $\mathcal{A}_{h_1 h_2 \to h_3 h_4}$ (defined for $s - 4m^2 \geq -t \geq 0$) can be analytically continued in the complex $s$-plane to a combination of $u$-channel matrix elements $\mathcal{A}_{h_1 \bar{h}_4 \to h_3 \bar{h}_2}$ (defined for $s  \leq 0 \leq  -t$). 
This means that, although in the region $s  \leq 0 \leq  -t$ the amplitude $\mathcal{A}_{h_1 h_2 h_3 h_4} (s,t)$ cannot be written as \eqref{eqn:Adef} or bounded by \eqref{eqn:unit_As}, we can instead use crossing to express it in terms of \eqref{eqn:Au_pw} and appeal to $u$-channel unitarity. 
 
However, this continuation from the $s$- to the $u$-channel effectively contains a Lorentz boost, and so unlike for scalar particles it is generally not true that $\mathcal{A}_{h_1 h_2 h_3 h_4} (s,t)$ and $\mathcal{A}_{h_1 \bar{h}_4 h_3 \bar{h}_2} (u,t)$ simply coincide, since they carry little group indices which are transformed by the Lorentz boost. 
The precise crossing relation between the $s$- and $u$-channel helicities amplitudes was worked out in detail in \cite{Trueman:1964zzb,cohen-tannoudji_kinematical_1968,Hara:1970gc,Hara:1971kj}. 
The final result is given in Appendix~\ref{app:crossing}, where we show that for elastic scattering (in which $S_3 = S_1$ and $S_4 = S_2$) the amplitude $A_{h_1 h_2 h_3 h_4} (s,t)$ can be expanded in $u$-channel partial waves, 
\begin{align}
\mathcal{A}_{h_1 h_2 h_3 h_4} (4m^2 -u -t , t)  &= \sum_{J_u} \sum_{h_a'}   \,  \langle J_u \, h_u^{\text{out} \, \prime} | e^{+i \hat{J}_y \theta_u} | J_u \, h_u^{\text{in} \, \prime} \rangle  \nonumber \\
&\times  C^{h_3 \bar{h}_2}_{h_3' \bar{h}_2'} (\chi_u) \; \langle h_3' \bar{h}_2' | \hat{T}  ( u, J_u , h_u^{\rm in \, \prime}  ) |  h_1' \bar{h}_4' \rangle \;
C_{h_1 \bar{h_4}}^{h_1' \bar{h}_4'} (\chi_u) \, , 
\label{eqn:Across_pw}
\end{align}
in the region $s \leq 0 \leq -t$, where the sum over $h_a'$ implies summing over all possible values of $h_1', \dots, h_4'$, with  $h_u^{\text{in} \, \prime} = h_1'-\bar{h}_4'$ and $h_u^{\text{out} \, \prime} =h_3'-\bar{h}_2' $.
Here, the crossing matrices $C_{h_1 \bar{h}_4}^{h_1' \bar{h}_4'} (\chi_u) $ are given by the matrix elements,
\begin{align}
C_{h_1 \bar{h}_4}^{h_1' \bar{h}_4'} (\chi_u) =   \langle h_1' \bar{h}_4' | e^{- i \hat{S}_y \chi_u } | h_1 \bar{h}_4 \rangle  \; , \;\;  C_{h_3' \bar{h}_2' }^{h_3 \bar{h}_2} (\chi_u) =  \langle h_3 \bar{h}_2 | e^{- i \hat{S}_y \chi_u } | h_3' \bar{h}_2' \rangle  \; ,
\label{eqn:Cdef}
\end{align}
where $\hat{S}_y | h_1 h_2 \rangle = \left( \hat{S}_y^{(1)} + \hat{S}_y^{(2)} \right) | h_1 h_2 \rangle$ implements a rest-frame rotation of each particle by the angle $\chi_u$ given in \eqref{eqn:chiu}, which is a result of the Lorentz boost required to go from $s$-channel kinematics \eqref{eqn:kinematics} to $u$-channel kinematics \eqref{eqn:kinematics_u}.

\paragraph{Forward limit:}
Now we can take the $\text{Abs}_u$ of \eqref{eqn:Across_pw}, and use unitarity of the $u$-channel process \eqref{eqn:Audef} to bound $\mathcal{A}_{h_1 h_2 h_3 h_4} ( s, t)$ for negative values of $s$. 
Again adopting $|T_u\rangle$ as a shorthand for the partial trace,
\begin{align}
\langle p_u \, J_u \, h_u |  \hat{T}^\dagger \hat{T} | p_u \, J_u \, h_u \rangle =  | T_u \rangle \langle T_u | \; , 
\end{align}
we see that taking the $\text{Abs}_u$ of \eqref{eqn:Across_pw} in the forward limit gives,
\begin{align}
& 2 \,  \text{Abs}_u \, \mathcal{A}_{h_1 h_2 h_1 h_2} (4m^2 -u  , 0 )   =  \sum_{J_u = | h_u |}^{\infty}    |  \langle T_u | h_1 \bar{h}_2 \rangle |^2 
\label{eqn:DiscAu_fwd}
\end{align}
for all $s \leq 0$, which is the crossing image of \eqref{eqn:DiscAs_fwd}.

\paragraph{First $\bm{t}$ derivative:}
However, when taking a $t$ derivative, this will act on the crossing matrices \eqref{eqn:Cdef}.
Explicitly, taking a $t$-derivative at fixed $u$ of \eqref{eqn:Across_pw} gives\footnote{
As with \eqref{eqn:dtDiscA}, \eqref{eqn:dt_DiscAu} should be understood as first differentiating and setting $t=0$, and then taking the absorptive part. 
}, 
\begin{align}
& (u- 4m^2) \partial_t \, \text{Abs}_u \, \mathcal{A}_{h_1 h_2 h_1 h_2} (4m^2 -u - t , t ) |_{t=0}  
\nonumber \\ 
 &=  \sum_{J_u = | h_u |}^{\infty}  \langle J_u h_u | \hat{J}_y^2 | J_u h_u \rangle  \,   | \langle T_u | h_1 \bar{h}_2 \rangle  |^2 + \frac{2m^2}{u} \langle \hat{S}_y^2 \rangle 
\label{eqn:dt_DiscAu}
\end{align}
where the $\langle \hat{S}_y^2 \rangle$ term comes from the crossing matrices (using that $\partial \cos \chi_u / \partial \cos \theta_u = m^2/u$ at $t=0$), and is given explicitly by,
\begin{align}
\langle \hat{S}_y^2 \rangle &=  \sum_{J_u = | h_u |}^{\infty}  \Big\{  \sum_{h_u'}   \langle T_u | \hat{P}_{h_u'} \hat{S}_y | h_1 \bar{h}_2 \rangle  \langle h_1 \bar{h}_2 | \hat{S}_y \hat{P}_{h_u'} | T_u \rangle    +  \text{Re} \left[  \langle T_u | h_1 \bar{h}_2 \rangle   \langle h_1 \bar{h}_2 | \hat{S}_y^2 \hat{P}_{h_u} | T_u \rangle      \right]   \Big\}   \; ,
\label{eqn:Sy2}
\end{align}
where we have defined the projection operator, 
\begin{align}
 \hat{P}_{h_u} = \sum_{ \substack{ h_1' , \, \bar{h}_2' \\ h_1' - \bar{h}_2' = h_u }} | h_1' \bar{h}_2' \rangle \langle h_1 ' \bar{h}_2' |  \; ,
\end{align}
which sums over all pairs of helicities with a fixed total $h_u$, and arises as a result of the $\delta_{h_u^{\text{out} \; \prime} \; h_u^{\text{in} \; \prime}}$ enforcing $J_z$-conservation when $\theta_u = 0$ in the $e^{+ i \hat{J}_y \theta_u}$ matrix element. 

While the first term in $\langle \hat{S}_y^2 \rangle$ is manifestly positive, the second term is not. This was pointed out (albeit in a different polarisation basis) in \cite{deRham:2017zjm}. 
There are two possible resolutions. Firstly, one could simply neglect the $\mathcal{O} (m^2)$ term in \eqref{eqn:dt_DiscAu}, making the implicit assumption that the UV completion is such that these terms are small (i.e. that no UV amplitude grows like $1/m^2$ at small $m$). This is the approach adopted in \cite{Remmen:2020uze} for the particular case of $S_1 = S_2 = 1/2$. 
However, we will see in explicit examples in Section~\ref{sec:examples} that for spins $\geq 1$ this massless must be taken with great care, since scattering of longitudinal modes can indeed scale like $1/m^2$ and lead to a violation of the crossing relation if the $\mathcal{O} (m^2)$ term is discarded prematurely.
The second resolution for the sign-indefinite terms in \eqref{eqn:dt_DiscAu}, and the strategy which we shall adopt in this work, is to consider a sum over all ingoing helicities with fixed total $h_u = h_1 - \bar{h}_2$,
\begin{align}
\mathcal{A}_{h_u} (s,t)  :=  \sum_{ \substack{ h_1 , \, h_2 \\ h_1 + h_2 = h_u }} \mathcal{A}_{h_1 h_2 h_1 h_2} (s,t)
\label{eqn:AhuDef}
\end{align}
Averaging over the relative helicity in this way produces a positive $\langle \hat{S}_y^2 \rangle$ in \eqref{eqn:dt_DiscAu},
\begin{align}
& (u- 4m^2) \partial_t \, \text{Abs}_u \, \mathcal{A}_{h_u} (4m^2 -u - t , t ) |_{t=0}  \nonumber \\ 
 &=  \sum_{J_u = |h_u|}^{\infty} \Bigg\{   
 \langle J_u h_u | \hat{J}_y^2 | J_u h_u \rangle \;   \left|  \hat{P}_{h_u} | T_u \rangle  \right|^2  
 + \frac{2 m^2}{u}  \left(  \sum_{h_u'}  \left|  \hat{P}_{h_u} \hat{S}_y \hat{P}_{h_u'}  | T_u \rangle  \right|^2  
 +  \left|   \hat{S}_y \hat{P}_{h_u}  | T_u \rangle  \right|^2 \right)  \Bigg\} \; .
\label{eqn:dt_DiscAhu}
\end{align}
Not only is the right-hand-side of \eqref{eqn:dt_DiscAhu} now manifestly positive, but in fact comparing with \eqref{eqn:Across_pw} and using the selection rule \eqref{eqn:Jy2_identity}, we find that, \vskip5pt
\begin{mdframed}[style=boxed] \vskip-15pt
\begin{align}
 (u- 4m^2) \partial_t \, \text{Abs}_u \, \mathcal{A}_{h_u} (4m^2 -u - t , t ) |_{t=0} 
 \geq
|h_u|  \text{Abs}_u \, \mathcal{A}_{h_u} (4m^2 -u  , 0 ) \; ,  
\label{eqn:unit_dtAu}
\end{align}
\end{mdframed}
as a consequence of $J_u \geq | h_u|$. This is the crossing image of the $s$-channel bound \eqref{eqn:unit_dtAs}.

~

Armed with the unitarity bounds \eqref{eqn:unit_dtAs} and \eqref{eqn:unit_dtAu}, which place lower bounds on the size of $\partial_t \text{Abs}_s \, \mathcal{A}_{h_1 h_2 h_1 h_2 } (s,0)$ and its crossing image which are stronger than the basic $\partial_t \text{Abs}_s \, \mathcal{A}_{h_1 h_2 h_1 h_2} (s,0) > 0$ condition, we can now derive stronger positivity bounds on the coefficients appearing in any low-energy EFT of massive spinning particles.

\section{Analyticity and a Stronger Positivity Bound}
\label{sec:positivity}

So far we have considered the consequences of the unitarity condition $\hat{S}^\dagger \hat{S} = \mathbbm{1}$ on the $2 \to 2$ scattering amplitude. 
However, in practice these amplitudes are computed using a low-energy effective description---since experimentally what we have access to is the behaviour of $\mathcal{A}$ at small values of $s$, this allows us to make theoretical predictions when the full UV theory (namely $\mathcal{A}$ at large $s$) is not explicitly known. 
 
We will now consider the question of whether a given low-energy EFT (i.e. a given $\mathcal{A}_{\rm EFT} (s)$ at small $s$) could have a UV completion (an analytic continuation to large $s$) which is compatible with these unitarity requirements. 
Demanding that there exists such a UV completion places constraints on the IR Wilson coefficients, known as positivity bounds.

\paragraph{Causality and analyticity:}
In order to relate the UV and IR behaviour of the amplitude, we will assume that the UV completion is \emph{causal}. Causality can be used to establish the analytic properties of the amplitude in the complex $s$-plane \cite{Bremermann:1958zz, bogoliubov1959introduction, Hepp_1964, Jin:1964zza, Martin:1965jj, Mahoux:1969um}, allowing one to leverage basic results from complex analysis. 
In particular, Cauchy's residue theorem allows us to evaluate $\mathcal{A}$ at a regular point $s_0$ using a contour integral which encircles $s_0$. By deforming this contour, $\mathcal{A}$ can be expressed as an integral over its discontinuities ($\text{Abs}$ parts), known as a dispersion relation. These dispersion relations connect the low-energy EFT amplitude (evaluated at some $s_0$ in the IR) to the value of $\text{Abs} \, \mathcal{A}$ in the UV, and are a standard ingredient in deriving positivity bounds. 

However, for massive spinning particles,  $\mathcal{A}_{h_1 h_2 h_1 h_2} (s,t)$ contains unphysical ``kinematic'' singularities at finite $t$, arising for instance from the factors of $\cos \theta_s/2$ and $\sin \theta_s/2$ used to define the polarisation tensors.  
These can also be seen in the partial wave expansion \eqref{eqn:Apw}, since the matrix elements $\langle J_s h_s^{\rm out} | e^{-i  \hat{J}_y \theta_s} | J_s h_s^{\rm in} \rangle$ contain poles and branch cuts when written in terms of $s$ and $t$. 
For example, supposing that $h_s = 1/2$ then the first term in \eqref{eqn:Apw} is proportional to
$ \langle \tfrac{1}{2} , \tfrac{1}{2} |  e^{- i \hat{J}_y \theta_s}  | \tfrac{1}{2} , \tfrac{1}{2} \rangle = \sqrt{ -su} / \sqrt{s-4m^2}$. This is single-valued in the physical $s$-channel region, but introduces a kinematic branch cut in the unphysical region that will contribute to any dispersion relation (and which is not constrained by unitarity). 

\paragraph{Regulated amplitude:}
In order to overcome this issue\footnote{
Another approach, described in \cite{Bellazzini:2016xrt}, is to choose this kinematic branch cut to be along the real axis interval $s \in [0,4m^2]$, which is certainly within the regime of validity of the EFT and can therefore be explicitly subtracted from any dispersion relation. 
}, we will consider a suitably regulated amplitude which is free from such unphysical singularities. 
These kinematic singularities were studied in detail in \cite{Cohen-Tannoudji:1968lnm}, where it was shown that for elastic processes in which the helicities are preserved, the \emph{regulated} amplitude 
\begin{align}
 \hat{\mathcal{A}}_{h_1 h_2 h_1 h_2} (s, t) := \mathcal{K} ( | h_s | ) \;  \mathcal{A}_{h_1 h_2 h_1 h_2} (s,t) \;\; \text{with} \;\; \mathcal{K} ( | h_s | ) := \frac{ \left(  s ( s - 4m^2  )  \right)^{S_1 + S_2}  }{ \left( - s u  \right)^{|  h_s |} }
 \label{eqn:Ahath1h2}
\end{align}
is free from any kinematic singularity. The only non-analyticities which appear in $ \hat{\mathcal{A}}_{h_1 h_2 h_1 h_2} (s, t)$ are those required by unitarity (and crossing). 

Note that removing the kinematic singularities in this way has introduced an additional $t$ dependence, and in fact,
\begin{align}
(s-4m^2)  \partial_t  \,  \hat{\mathcal{A}}_{h_1 h_2 h_1 h_2} (s, t) |_{t=0} = \mathcal{K} (| h_s |)  \left[ (s-4m^2) \partial_t - | h_s |  \right]  \,  \mathcal{A}_{h_1 h_2 h_1 h_2} (s, t) |_{t=0}
\end{align}
and the $s$-channel unitarity bound \eqref{eqn:unit_dtAs} becomes simply,
\begin{align}
 \partial_t \, \text{Abs}_s \,  \hat{\mathcal{A}}_{h_1 h_2 h_1 h_2} (s, t ) |_{t=0} \geq 0\,  ,
\label{eqn:unit_dtAs_hat}
 \end{align}
in the region $s-4m^2 \geq 0$. Although this superficially resembles the usual positivity condition $\partial_t \text{Abs}_s \, \mathcal{A}_{h_1 h_2 h_1 h_2} |_{t=0} > 0$, note that without our stronger condition \eqref{eqn:unit_dtAs} we would not have been able to remove the factor of $(-su)^{|h_s|}$ in \eqref{eqn:Ahath1h2} and retain positivity of the $s$-channel branch cut. 

For the $u$-channel unitarity bound \eqref{eqn:unit_dtAu}, recall that it was necessary to sum over all pairs of helicities $(h_1,h_2)$ with fixed $h_1 + h_2 = h_u$. This sum will have a minimum value of $|h_1 - h_2|$, which we denote $|h_s|_{\rm min}$. The procedure analogous to \eqref{eqn:Ahath1h2} for removing kinematic singularities is therefore to define the regulated amplitude,
\begin{align}
 \hat{\mathcal{A}}_{h_u} (s, t) := \mathcal{K} ( | h_s |_{\rm min} ) \;  \mathcal{A}_{h_u} (s,t)  \; , 
 \label{eqn:Ahat}
\end{align}
where the kinematic factor $\mathcal{K}$ is defined in \eqref{eqn:Ahath1h2}.
Unitarity in the $u$-channel \eqref{eqn:unit_dtAu} therefore implies,
\begin{align}
&  \partial_t \, \text{Abs}_u  \, \hat{\mathcal{A}}_{h_u} ( 4m^2 -u -t, t) |_{t=0}  \nonumber \\
&\geq \left(  \frac{ |h_u| - |h_s|_{\rm min} }{u-4m^2} +  \left(  S_1 +  S_2 \right) \left( \frac{1}{u-4m^2} + \frac{1}{u}   \right)  \right) \text{Abs}_u  \, \hat{ \mathcal{A} }_{h_u} (4m^2-u, 0) \; , 
\label{eqn:unit_dtAu_hat}
\end{align}
in the region $s \leq 0$.

\begin{figure}[t!]
\centering
\includegraphics[width=0.65\textwidth]{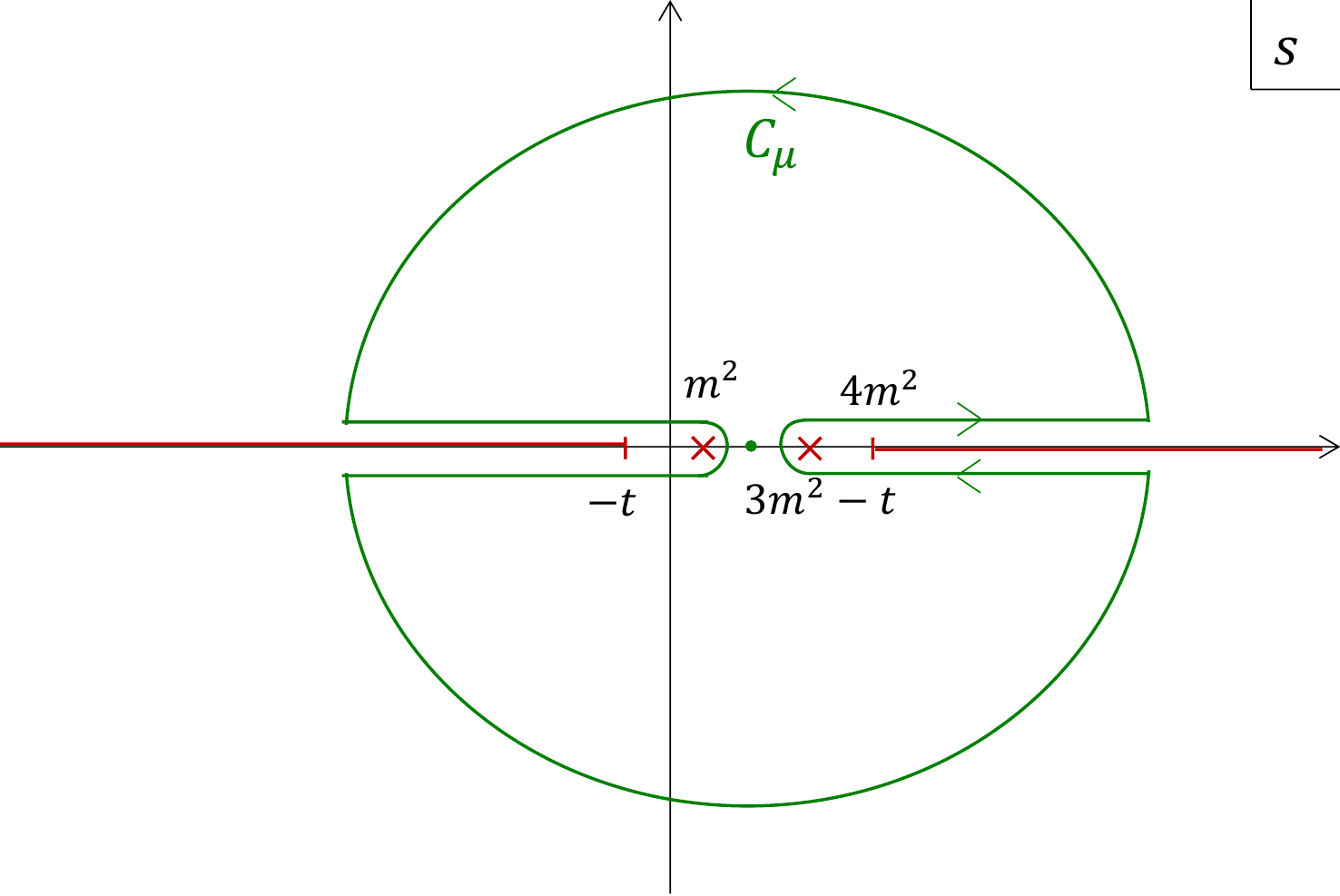}
\caption{The contour $C_\mu$ used to define $\hat{\mathcal{A}}^{(2N)}_{h_u}$ in \eqref{eqn:An}.}
\label{fig:contour}
\end{figure}

\paragraph{Dispersion relations:}
We can now use the analyticity of $\hat{\mathcal{A}}_{h_u} (s , t)$ in the complex $s$-plane to derive a dispersion relation at fixed $t$. 
Following the by now standard procedure \cite{Adams:2006sv}, we will consider the contour integrals\footnote{
The most general fixed $t$ dispersion relation would integrate $\hat{\mathcal{A}}$ against a general integration kernel $F(s)$ which is analytic for all $|s| \geq \Lambda$, the EFT cut-off. 
In this high-energy regime, $F(s)$ can therefore be expanded in a Laurent series, and so in general we have a linear combination of the integrals \eqref{eqn:An}.
}, 
\begin{align}
&\hat{\mathcal{A}}^{(2N)}_{h_u} ( s_0, t ; \mu ) := \oint_{ C_\mu } \frac{d s}{2 \pi i} \frac{  
\hat{\mathcal{A}}_{h_u} (s,t) }{ ( s - s_0 )^{2N+1} }  \; , 
\label{eqn:An}
\end{align}
where $C_\mu$ is a circular contour of radius $\mu$ centered at the crossing-symmetric point $s=2m^2-t/2$ and $s_0$ is a point within this region. 
By Cauchy's theorem, $\hat{\mathcal{A}}^{(2N)}_{h_u} ( s_0, t ; m^2 - t/2 ) = \partial_s^{2N} \hat{\mathcal{A}}_{h_u} ( s_0, t)$ for any $s_0$ in the region $m^2 < s_0 < 3m^2-t$, and can be computed within the low-energy EFT. 
When $\mu$ is greater than $m^2-t/2$, the contour integral in \eqref{eqn:An} is to be understood as a deformation of the $C_{m^2-t/2}$ contour on the physical sheet, as shown in Figure~\ref{fig:contour}. 
The observable $\hat{\mathcal{A}}^{(2N)}_{h_u} ( s_0, t ; \mu )$ therefore corresponds to $\partial_s^{2N}$ of the amplitude evaluated at $s_0$ with all poles and branch cuts subtracted up to the scale $\mu$.
Changing the radius $\mu$ of the contour integral shifts $\hat{A}^{(2N)}_{h_u} (s_0,t; \mu)$ by the real axis branch cut discontinuities, and can be interpreted as integrating in/out physics at that scale\footnote{
These $\hat{\mathcal{A}}^{(2N)}_{h_u} ( s_0, t ; \mu )$ are in many ways analogous to the arcs defined in \cite{Bellazzini:2020cot}, but note that since crossing symmetry is not trivial for massive spinning particles one can no longer use an arc in the upper half-plane only since the resulting dispersion relation contains $\mathcal{A}_{h_1 h_2 h_3 h_4} (s+i\epsilon,t)$ and $\mathcal{A}_{h_1 \bar{h}_4 h_3 \bar{h}_2} (s - i \epsilon , t)$ only (which cannot be combined into an $\text{Abs}$ part and thence constrained by unitarity). 
}. 
Explicitly\footnote{
Note that with our conventions $\text{Abs}_s$ is related to $\text{Abs}_u$ by a minus sign, namely $\text{Abs}_s \, \mathcal{A}_{h_1 h_2 h_3 h_4} ( 4m^2 - u - t , t ) = - \text{Abs}_u \, \mathcal{A}_{h_1 h_2 h_3 h_4} ( 4m^2 - u - t , t )$. 
}, 
\begin{align}
&\hat{\mathcal{A}}^{(2N)}_{h_u} ( s_0, t ; \mu_{\rm IR} ) - \hat{\mathcal{A}}^{(2N)}_{h_u} ( s_0, t ; \mu_{\rm UV} ) 
\nonumber \\
&= 
 \int_{\mu_{\rm IR}}^{\mu_{\rm UV}} \frac{ds}{\pi} \frac{ \text{Abs}_s \, \hat{\mathcal{A}}_{h_u} (s,t) }{ (s-s_0)^{2N+1}}  
+ \int_{\mu_{\rm IR}}^{\mu_{\rm UV}} \frac{du}{\pi} \frac{ \text{Abs}_u \, \hat{\mathcal{A}}_{h_u} (4m^2-u-t,t) }{ (u-u_0)^{2N+1}} \; , 
\label{eqn:An_disp}
\end{align}
where $u_0 = 4m^2-s_0-t$. 
The dispersion relation \eqref{eqn:An_disp} can be used to relate an IR scale $\mu_{\rm IR}$, at which the EFT can be used to compute $\hat{\mathcal{A}}_{h_u}^{(2N)}$, to a UV scale $\mu_{\rm UV}$, at which $\hat{\mathcal{A}}_{h_u}^{(2N)}$ is sensitive to certain properties of the UV completion. 

\paragraph{Forward limit:}
Since unitarity requires that both of the discontinuities in \eqref{eqn:An_disp} are positive in the forward limit (see \eqref{eqn:DiscAs_fwd} and \eqref{eqn:DiscAu_fwd}), this establishes that $\hat{\mathcal{A}}^{(2N)}_{h_u} ( s_0, 0 ; \mu )$ is a \emph{monotonic} function of the scale $\mu$. 
In particular\footnote{
In fact, since $\text{Abs}_u \, \mathcal{A}_{h_1 \bar{h}_2 h_1 \bar{h}_2} (s,0) > 0$ in the forward limit, there is no need to perform the helicity sum \eqref{eqn:AhuDef} since every elastic $\mathcal{A}_{h_1 h_2 h_1 h_2}^{(2N)} (s_0, 0 ; \mu )$ is monotonic~\cite{Bellazzini:2016xrt}. In fact, the elastic scattering of any superposition of helicity states is also monotonic~\cite{Cheung:2016wjt}, and more recently bounds have also been placed on the inelastic configurations~\cite{Zhang:2020jyn, Trott:2020ebl}.  
},
\begin{align}
 \hat{\mathcal{A}}^{(2N)}_{h_u} ( s_0, 0 ; \mu_{\rm IR} ) > 
\hat{\mathcal{A}}^{(2N)}_{h_u} ( s_0, 0 ; \infty ) \; ,
\label{eqn:muIR}
\end{align}
for all $4m^2 - \mu_{\rm IR} < s_0 < \mu_{\rm IR}$, where $\mu_{\rm IR}$ is any low-energy scale which is resolved by the EFT. 
This allows us to bound EFT coefficients in terms of the high-energy behaviour $\hat{\mathcal{A}}^{(2N)}_{h_u} ( s_0, 0 ; \infty )$. 
Sum rules like \eqref{eqn:muIR} connect low-energy EFT observable to properties of the UV physics, and become fully-fledged positivity bounds when $\hat{\mathcal{A}}^{(2N)}_{h_u} ( s_0, t ; \infty )$ vanishes.

Taking the $t$-derivative of \eqref{eqn:An_disp} will similarly produce a bound on $\partial_t  \hat{\mathcal{A}}^{(2N)}_{h_u} ( s_0, t ; \mu_{\rm IR} )|_{t=0}$ in terms of $\partial_t  \hat{\mathcal{A}}^{(2N)}_{h_u} ( s_0, t ; \infty )|_{t=0}$. 
Before we describe this $t$-derivative bound explicitly, let us briefly describe the conditions under which the quantities $\hat{\mathcal{A}}^{(2N)}_{h_u} ( s_0, t ; \infty )$ vanish.

\paragraph{Locality and convergence:}
The standard assumptions of unitarity, causality and locality in any gapped quantum field theory lead to the Froissart-Martin bound \cite{Froissart:1961ux, Martin:1962rt}, 
\begin{align}
  \lim_{|s|\to \infty} | \mathcal{A}_{h_1 h_2 h_1 h_2} (s, t) | < s^2 \; , 
 \label{eqn:Froissart}
\end{align}
which requires that $\hat{\mathcal{A}}^{(2N)}_{h_u} ( s_0, 0 ; \infty )$ vanishes for sufficiently large $N$. 
Notice that the normalisation used to regulate the amplitude in \eqref{eqn:Ahat} depends on $s$ like $\mathcal{K} ( |h_s|_{\rm min} ) \sim s^{2 ( S_1 +  S_2 - |h_s|_{\rm min} )}$, and so it is convenient to define,
\begin{align}
\hat{N}:= N - S_1 -  S_2 +  | h_s |_{\rm min}  \, ,
\end{align}
since then,
\begin{align}
\hat{\mathcal{A}}^{(2N)}_{h_u} ( s_0, 0 ; \infty ) \sim \int_{C_\infty} \frac{ds}{s} \;  \frac{ \mathcal{A}_{h_u} (s, 0) }{ s^{2 \hat{N}} } = 0 \;\; \text{for} \;\; \hat{N}  \geq 1 
\label{eqn:Froissart2}
\end{align}
in any local (unitary, causal) UV completion. In other words, the number $\hat{N}$ can be thought of as (half) the number of ``spin-adjusted'' derivatives acting on the amplitude.

\begin{table}[t]
\begin{center}
\begin{tabular}{|c|c|c|c|c|c|}
\hline
&  $N_{UV}$& $\hat{\mathcal{A}}^{(2N)} > 0$ bound if & $\partial_t \hat{\mathcal{A}}^{(2N)}$ bound if & Example UV  \\
\hline
Super-Froissart  & 1/2 & $\hat{N} \geq 1$ & $\hat{N} \geq 0$ & No $t$-channel \\
Froissart  & 1 & $\hat{N} \geq 1$ & $\hat{N} \geq 1$ & Any local QFT \\
Sub-Froissart  & 3/2 & $\hat{N} \geq 2$ & $\hat{N} \geq 1$ & Galileon? \\
\hline
\end{tabular}
\end{center}
\caption{Summary of three different high-energy growths \eqref{eqn:NUVdef} of the amplitude in the UV and the corresponding positivity bounds.}
\label{tab:convergence}
\end{table}

From \eqref{eqn:muIR}, we therefore conclude that $ \hat{\mathcal{A}}^{(2N)}_{h_u} ( s_0, 0 ; \mu_{\rm IR} )$, which equals $\partial_s^{2N}  \hat{\mathcal{A}}_{h_u} (s_0, 0)$ minus the EFT branch cut contributions out to the scale $\mu_{\rm IR}$, must be positive for all  $\hat{N} \geq 1$, i.e. for all $N \geq 1 + S_1 + S_2 - |h_s|_{\rm min}$, if a local, unitary, and causal UV completion is to exist \cite{deRham:2017zjm}. This is the analogue of the simple bound $\partial_s^2 \mathcal{A} > 0$ that is well known for scalar amplitudes \cite{Adams:2006sv}.
We emphasize that it is the spin-dependent number $\hat{N}$ that determines the operator dimension at which these bounds first become constraining. For instance, this forward limit bound constrains operators of dimension $4 + 4 \hat{N}$.  

In addition to Froissart-boundedness \eqref{eqn:Froissart}, there are various other asymptotic behaviour of the amplitude at high $s$ that one could consider (see e.g. \cite{Tokuda:2019nqb}). 
In many classes of UV completion the convergence at large $s$ is faster than the Froissart bound; in string theory for instance, the amplitude is often exponentially bounded at large $s$, and so $\hat{\mathcal{A}}^{(2N)}_{h_u} ( s_0, 0 ; \infty ) = 0$ for any $\hat{N} \geq 0$. 
Rather than considering only the set of Froissart-bounded UV theories \eqref{eqn:Froissart}, we can also consider other classes of UV theories whose amplitudes are bounded by a different asymptotic growth rate, specified by,
\begin{align}
\lim_{s\to \infty} | \mathcal{A}_{h_1 h_2 h_1 h_2} (s,t) | < s^{2N_{\rm UV}} \; .
\label{eqn:NUVdef}
\end{align}
For these  more (or less) restrictive UV completions, $ \hat{\mathcal{A}}^{(2N)}_{h_u} ( s_0, 0 ; \mu_{\rm IR} ) > 0$ for all integer $\hat{N} \geq N_{\rm UV}$. 

One particular class of UV completion which we consider below are those in which the amplitude is softer than $p^2$ at large momentum, i.e. \eqref{eqn:NUVdef} with growth $N_{\rm UV} = 1/2$ (where we are not distinguishing between $t$ and $s$).  
This high-energy behaviour has been discussed also in \cite{Bellazzini:2016xrt, Remmen:2020uze, Gu:2020thj}, and includes for instance tree-level UV completions with no $t$-channel exchange. Such a ``super-Froissart'' convergence would imply that,
\begin{align}
\text{Super-Froissart:} \;\;\;\; \partial_t \hat{\mathcal{A}}^{(2N)}_{h_u} ( s_0, t ; \infty ) |_{t=0} = 0  \;\; \text{for} \;\; \hat{N} \geq 0  \; . 
\label{eqn:superFroissart2}
\end{align}
Let us stress that unlike \eqref{eqn:Froissart2}, which is a robust requirement of locality, this condition \eqref{eqn:superFroissart2} is simply a restriction on the kinds of UV completion for which we are allowing. 

Finally, there are also cases in which one may want to allow for a sub-Froissart growth in the UV---most notably for the Galileon interactions \cite{Nicolis:2008in}, which violate the bound $\mathcal{A}^{(2)} > 0$ and so may require a UV completion which violates~\eqref{eqn:Froissart}  (see e.g. \cite{Keltner:2015xda}). In this case, by considering \eqref{eqn:NUVdef} with $N_{\rm UV} = 3/2$, we can allow for some mild non-locality (i.e. a violation of the Froissart bound by just one power of $s$). This implies, 
\begin{align}
\text{Sub-Froissart:}  \;\;\;\; \partial_t \hat{\mathcal{A}}^{(2N)}_{h_u} ( s_0, t ; \infty ) |_{t=0} = 0  \;\; \text{for} \;\; \hat{N} \geq 1  \; . 
\label{eqn:subFroissart2}
\end{align}
These different possible UV behaviours are summarised in Table~\ref{tab:convergence}.

\paragraph{First ${\bm t}$ derivative:}
With all of the ingredients now in place, we will complete the derivation of positivity bounds for the first $t$ derivative of a general elastic $2\to 2$ amplitude $\mathcal{A}_{h_1h_2h_1h_2}$ for massive particles with arbitrary spins, which follow from unitarity, causality and locality (or a super/sub Froissart-boundedness) in the UV. 
Taking the $t$ derivative of the dispersion relation \eqref{eqn:An_disp}, evaluated at $s_0 = 4m^2$, \begin{align}
\partial_t \hat{\mathcal{A}}_{h_u}^{(2N)} ( 4m^2 ,t ; \mu ) |_{t=0}  = 
 \partial_t \hat{\mathcal{A}}_{h_u}^{(2N)} ( 4m^2  ,t ; \infty ) |_{t=0} 
+ \int_{\mu}^{\infty} \frac{ds}{\pi} \frac{I_s }{ (s-4m^2)^{2N+1}}  
+
\int_{\mu}^{\infty} \frac{du}{\pi} \frac{I_u}{ u^{2N+1} }
\label{eqn:dt_disp}
\end{align}
where the $s$-channel contribution is positive thanks to the unitarity bound \eqref{eqn:unit_dtAs_hat},
\begin{align}
I_s  = \partial_t \, \text{Abs}_s \, \hat{\mathcal{A}}_{h_u} (s,t) \geq 0  \; ,
\label{eqn:Is}
\end{align}
and the $u$-channel contribution is bounded by,
\begin{align}
I_u &= \partial_t \, \text{Abs}_u \, \hat{\mathcal{A}}_{h_u} (4m^2-u-t,t) |_{t=0}  - \frac{2N+1}{u}  \text{Abs}_u \, \hat{\mathcal{A}}_{h_u} (4m^2-u,0)  \nonumber \\
&> \frac{1}{u} \left( | h_u | + | h_s |_{\rm min}  - 2 \hat{N} - 1   \right) \text{Abs}_u \, \hat{\mathcal{A}}_{h_u} (4m^2-u , 0) \; , 
\label{eqn:Iu}
\end{align}
thanks to the unitarity bound \eqref{eqn:unit_dtAu_hat} (using that $S_1 +S_2 \geq |h_s|_{\rm min}$ and $1/(u-4m^2) > 1/u$ for all $u > \mu \geq 4m^2$).
We introduce a new variable for this particular combination of the helicities and spins,
\begin{align}
\alpha := 2 \hat{N} + 1 - |h_u| -  | h_s|_{\rm min} \; , 
\end{align}
so that $I_u > 0$ if $\alpha <0$ and $I_u < 0$ if $\alpha > 0$. 
We will now discuss each case in turn. 

\paragraph{Small helicities ($\bm{\alpha > 0}$):}
When $\alpha$ is positive, the $s$- and $u$-channel branch cuts in \eqref{eqn:dt_disp} have opposite signs. This is the same issue encountered for scalar amplitudes (for which $\alpha$ is always positive). One resolution, first described in \cite{deRham:2017avq}, is to add to $\partial_t \hat{\mathcal{A}}^{(2N)} |_{t=0}$ enough of the positive $\hat{\mathcal{A}}^{(2N)} |_{t=0}$ to correct for the negative $I_u$. In our case, this allows us to establish the analogue of $\mu$ monotonicity \eqref{eqn:muIR} for $t$ derivatives,
\begin{align}
\left( \partial_t + \frac{\alpha}{\mu} \right) \hat{\mathcal{A}}_{h_u}^{(2N)} ( 4m^2 ,t ; \mu ) |_{t=0} 
>
\left( \partial_t + \frac{\alpha}{\mu} \right) \hat{\mathcal{A}}_{h_u}^{(2N)} ( 4m^2 ,t ; \infty ) |_{t=0} \; .
\end{align} 
We therefore arrive at the following positivity bound, where the range of $N$ is determined by which UV growth~\eqref{eqn:NUVdef} is assumed,\vskip5pt
\begin{mdframed}[style=boxed]  \vskip-15pt
\begin{align}
\left( \partial_t + \frac{\alpha}{\mu} \right) \hat{\mathcal{A}}_{h_u}^{(2N)} ( 4m^2 ,t ; \mu ) |_{t=0}  > 0 \;\; \text{for}  \;\;  
\hat{N} \geq \mathrm{max} \left(N_{\rm UV} ,\frac{|h_u| + |h_s|_{\rm min} -1}{2} \right)  \; .
\label{eqn:pos_main1}
\end{align}
\end{mdframed}
When $|h_u| = |h_s|_{\rm min} = 0$, this reproduces the scalar positivity bound of \cite{deRham:2017avq}, and when rotated to the transversity basis this represents a stronger version of the bounds in \cite{deRham:2017zjm} (which effectively used an $\alpha_{\rm there} = 2N + 1$, which is always $\geq \alpha_{\rm here}$).

\paragraph{Large helicities ($\bm{\alpha \leq 0}$):}
The main breakthrough achieved by our stronger unitarity condition is that when $\alpha \leq 0$ (i.e. for sufficiently large total helicity), $I_u > 0$ and $\partial_t \hat{\mathcal{A}}_{h_u}^{(2N)}|_{t=0}$ alone is a monotonic function of $\mu$, and consequently we have a more powerful bound,  \vskip5pt
\begin{mdframed}[style=boxed]  \vskip-15pt
\begin{align}
 \partial_t  \hat{\mathcal{A}}_{h_u}^{(2N)} (4m^2 , t ; \mu ) |_{t=0} > 0   \;\; \text{for}  \;\;  
 \frac{|h_u| + |h_s|_{\rm min} -1}{2} \geq   \hat{N}   \geq  N_{\rm UV} - \frac{1}{2}   \; . 
 \label{eqn:pos_main2}
\end{align}
\end{mdframed}
This is qualitatively stronger than \eqref{eqn:pos_main1} because with no $\mu^{-1} \hat{\mathcal{A}}_{h_u}^{(2N)} (4m^2 , t ; \mu )$ term, the bound \eqref{eqn:pos_main2}:
\begin{itemize}

\item[(i)] does not require a large $\mu$ to be useful (in contrast, when the EFT resolves only small values of $\mu$, then \eqref{eqn:pos_main1} simply reproduces the forward limit bound $\hat{\mathcal{A}}^{(2N)}_{h_u} ( 4m^2 , 0 ; \mu ) > 0$),

\item[(ii)] can be used to constrain dimension-6 operators, assuming super-Froissart convergence \eqref{eqn:superFroissart2} in the UV (in contrast, \eqref{eqn:pos_main1} only applies to $\hat{N} = 0$ when one assumes an even stronger convergence of $\mathcal{A}$ at large $s$, namely $N_{\rm UV} = 0$), 

\item[(iii)] can be used to constrain dimension-10 operators (Galileon-like interactions) even when the UV amplitude convergences slower than Froissart \eqref{eqn:subFroissart2}---of course for scattering scalar fields $\alpha$ is always positive and \eqref{eqn:pos_main1} applies, but in various ``IR completions'', such as Proca or massive gravity, the scattering of spinning particles can be constrained by \eqref{eqn:pos_main2} (see e.g. \cite{Cheung:2016yqr, Bonifacio:2016wcb, Bellazzini:2017fep, deRham:2017xox, deRham:2018qqo, Alberte:2019xfh, Alberte:2019zhd} for earlier positivity bounds in these theories).

\end{itemize} 

Finally, note that the sum over relative helicity was only necessary to remove the $\mathcal{O} (m^2)$ correction to the crossing relation, and so more generally one may write, 
\begin{align}
 \partial_t  \hat{A}_{h_1 h_2 h_1 h_2}^{(2N)} (s, t ; \mu) |_{t=0} > \mathcal{O} \left( \frac{m^2}{\mu}  \langle \hat{S}_y^2 \rangle  \right)   \;\; \text{when} \;\; \frac{|h_u| + |h_s| -1}{2} \geq   \hat{N}  \geq  N_{\rm UV} - \frac{1}{2}   \; ,
 \label{eqn:pos_main3}
\end{align}
where $\langle \hat{S}_y^2 \rangle$ is given in \eqref{eqn:Sy2}. 
Making an additional assumption about the UV to discard this term when $\mu \gg m^2$ (namely that this matrix element grows slower than $\sim \mu / m^2$ at small masses), then one can place a positivity bound directly on $\partial_t \partial_s^{2N} \hat{A}_{h_1 h_2 h_1 h_2}$ without the need for the sum over $h_s$ at fixed $h_u$.

Our bounds \eqref{eqn:pos_main1} and \eqref{eqn:pos_main2} (and \eqref{eqn:pos_main3}) are very general results: together they can be used to constrain the scattering of massive particles with \emph{any} spin and helicity. 
Note that for the special case $S_1 = S_2 = 1/2$ and $h_s = 1$, $h_u = 0$ recently considered in \cite{Remmen:2020uze}, we have that,
\begin{align}
 \hat{A}_{+-+-} (s,t) = \frac{ s-4m^2 }{ s - 4m^2  + t }  \; \mathcal{A}_{+-+-} (s,t) \; , 
\end{align}
and so the massless limit of the bound \eqref{eqn:pos_main2} with $\hat{N} = 0$ derivatives (assuming the super-Froissart growth \eqref{eqn:superFroissart2}) becomes,
\begin{align}
\lim_{m \to 0}  \partial_t  \hat{A}_{+-+-} (s, t) |_{ \substack{ s=4m^2 \\ t=0 }} 
= \left( \partial_t \mathcal{A}_{+-+-}  - \partial_s \mathcal{A}_{+-+-} \right) |_{\substack{s=0 \\ t =0}} \; , 
\label{eqn:RR}
\end{align}
which coincides precisely with the dimension-6 bound derived in \cite{Remmen:2020uze}.

\section{Some Applications}
\label{sec:examples}

In this section, we present some explicit examples of various spins and helicities and show how our positivity bounds lead to new and qualitatively stronger constraints on the coefficients appearing in the low-energy EFTs of massive spinning particles.

\subsection{Spinor-Scalar Scattering}

Let us begin by considering the following dimension-7 interaction,
\begin{align}
\mathcal{L}_{\rm int} = \tfrac{1}{2} \lambda_7 ( \partial \phi )^2 \bar \psi \psi \; , 
\label{eqn:Lint_1}
\end{align}
which couples a massive scalar field $\phi$ to a massive Dirac spinor $\psi$. 
For instance, this is the leading-order interaction compatible with an approximate shift symmetry for $\phi$ (which is softly broken by its mass). Modulo $SU(2)_L$ gauge indices and covariant derivatives, a similar interaction appears in the SMEFT as a Higgs-lepton coupling (see e.g. $\mathcal{O}_1$ in \cite{Bhattacharya:2015vja}), where it can lead to charged lepton number violation interactions with $W$ bosons, radiatively generate Majorana masses for neutrinos, and contribute to neutrinoless double beta decay (albeit at subleading order \cite{delAguila:2012nu}). 

We have chosen to give \eqref{eqn:Lint_1} as our first example because the $2 \to 2$ scattering amplitude $\psi \phi \to \psi \phi$ is particularly simple, 
\begin{align}
 \mathcal{A}_{\pm0\pm0}^{\psi \phi \to \psi \phi} (s,t) = \lambda_7 m ( t - 2m^2) \frac{ \sqrt{-s u} }{\sqrt{s (s-4m^2)}}  \; . 
\end{align}
We see that $\mathcal{A}$ indeed has the kinematic singularities described in Section~\ref{sec:positivity}, which can be removed by multiplying by the factor in \eqref{eqn:Ahat}, giving a regulated amplitude,
\begin{align}
\hat{\mathcal{A}}_{\pm}^{\psi \phi \to \psi \phi} (s,t) = 
\frac{\sqrt{s (s-4m^2)}}{\sqrt{-s u}} \mathcal{A}_{\pm0\pm0}^{\psi \phi \to \psi \phi} (s,t)
 = \lambda_7 m ( t - 2 m^2) \; ,
\end{align}
which is now analytic in $s$ and $t$ (as expected of a tree-level scattering amplitude). 

Furthermore, note that since this interaction has mass dimension $<8$, the forward-limit positivity bound $\mathcal{A}^{(2)} > 0$ cannot constrain $\lambda_7$ (assuming a local, Froissart-bounded, UV completion). 
However, following the example of \cite{Remmen:2020uze, Gu:2020thj} and assuming a super-Froissart boundedness \eqref{eqn:superFroissart2}, we can apply the positivity bound \eqref{eqn:pos_main2} with $N = 0$ $s$-derivatives (since $\alpha = 2 N$ because $|h_u| + |h_s|_{\rm min} = 1$ and $\hat{N} = N$ in this example). This gives the bound,
\begin{align}
\partial_t \hat{\mathcal{A}}_{\pm}^{\psi \phi \to \psi \phi}  =  \lambda_7 m > 0 \;\;\;\; \left( \text{when } N_{\rm UV} = 1/2 \right) . 
\end{align}
Physically, this means that such an interaction can only have $\lambda_7 < 0$ if $\partial_t \mathcal{A} \neq 0$ at high energies (for instance because the UV theory contains the $t$-channel exchange of a new heavy state). 
The sign of $\lambda_7$ is therefore a useful diagnostic of this property of the underlying UV completion.

\subsection{Vector-Scalar Scattering}

Next, we consider some simple interactions between a massive vector field, $A_\mu$, and a massive scalar field, $\phi$. This will help illustrate how the longitudinal modes of a massive vector field can lead to a violation of the crossing relation, and strong coupling at a parametrically lower scale, unless the massless limit is taken carefully.

\paragraph{Leading order:}
The leading-order interaction that mediates $A_\mu \phi \to A_\mu \phi$ is,
\begin{align} 
\mathcal{L}_{\rm int} = \tfrac{1}{4} \lambda_6 A^\mu A_{\mu} \phi^2 \; ,
\label{eqn:VS_dim6}
\end{align}
and the corresponding amplitude is simply,
\begin{align}
 \mathcal{A}_{h_1 0 h_3 0} (s,t) = \lambda_6 \epsilon_{h_1} (\bfp_1) \cdot \epsilon_{h_3}^* (\bfp_3) 
\end{align}
where the polarisation tensors $\epsilon_{h_1}^\mu (\bfp_1)$ for $A_\mu$ are given explicitly in Appendix~\ref{app:polarisation}, and are contracted using $\eta_{\mu\nu} = \text{diag} \left( -1, +1,+1,+1\right)$. 
For the transverse polarisations,
\begin{align}
 \mathcal{A}_{\pm 0 \pm 0} (s,t) = \lambda_6  \frac{-u}{s-4m^2}  \;\; , \;\;\;\;
  \mathcal{A}_{ \pm 0 \mp 0} (s,t) = \lambda_6 \frac{-t}{s-4m^2} \; . 
  \label{eqn:VS_trans}
\end{align}
Note the presence of the kinematic singularity at $s=4m^2$ when $t \neq 0$, and in particular that $(s-4m^2) \partial_t \mathcal{A}_{+0+0} |_{t=0} = 1$ is finite as $s \to 4m^2$ and the unitarity bound \eqref{eqn:unit_dtAs} is satisfied for any $\lambda$.  
For the longitudinal polarisation on the other hand,
\begin{align}
 \mathcal{A}_{0000} (s,t) = \lambda_6  \left(  1 + \frac{s}{2m^2} \,  \frac{ t }{ s - 4 m^2}  \right) \; ,
 \label{eqn:VS_long}
\end{align}
and $(s-4m^2) \partial_t \mathcal{A}_{0000} |_{t=0} = \lambda_6 s/2m^2$ so unitarity is violated at $s = 2m^2/\lambda_6$, indicating that $\lambda_6 \sim \mathcal{O} (m^2/\Lambda^2)$ is required if the EFT is to remain perturbative up to a cut-off scale $\Lambda$. This can also be seen by introducing Stuckelberg fields, $A_\mu \sim \partial_\mu \pi/m$, which makes manifest that $\lambda_6 A^\mu A_\mu \phi^2$ is a dimension-6 operator which becomes strongly coupled at the scale $m^2/ \lambda_6$. 

The kinematic singularities in \eqref{eqn:VS_trans} and \eqref{eqn:VS_long} disappear when $t=0$, but at finite $t$ they must be removed by \eqref{eqn:Ahath1h2} in order to apply a dispersion relation, 
\begin{align}
 \hat{\mathcal{A}}_{\pm 0 \pm 0} (s,t) = \lambda_6  \;\; , \;\;\;\;
  \hat{\mathcal{A}}_{ 0 0 0 0} (s,t) = \lambda_6  \frac{s}{2m^2} \left(  s t + 2 m^2 s - 8 m^4 \right)   \; . 
\end{align}
These cannot be constrained by positivity without assuming a rather strong convergence in the UV, namely $N_{\rm UV} = 0$. 
However, they serve to highlight an important point about the crossing relation. If the massless limit is taken with $t \neq 0$, then the resulting amplitudes do \emph{not} obey a trivial crossing relation, 
\begin{align}
\lim_{m \to 0} \mathcal{A}_{+0+0} (u,t) &= \lambda_6 \frac{s}{s + t} \neq \lim_{m\to 0}   \mathcal{A}_{+0+0} (s,t)   \nonumber \\
\lim_{m \to 0}  \mathcal{A}_{+0-0} (u,t) &= \lambda_6 \frac{t}{s+t} \neq  \lim_{m\to 0} \mathcal{A}_{+0-0} (s,t)  \; . 
\end{align}
The reason for this is that, while the crossing matrix $C_{h_1 \bar{h}_4}^{h_1' h_2'} (\chi_u) = \delta_{h_1}^{h_1'} \delta_{\bar{h}_4}^{h_2'} + \mathcal{O} (m)$, the scattering of longitudinal polarisations grows at small $m$, for instance $\mathcal{A}_{0000} \sim \mathcal{O} \left( 1/m^2 \right)$, and so one must consistently keep the subleading terms in the crossing matrix when expanding in small masses\footnote{
Alternatively, taking $m \to 0$ with $\Lambda = m/ \lambda_6$ held fixed effectively sends $\lambda_6 \to 0$ and removes this operator entirely in the massless limit, so a sufficiently careful power counting would ensure that all longitudinal interactions decouple as $m\to 0$.  
}.

\paragraph{Next-to-leading order:}
At next-to-leading order, there are various (dimension-8) interactions of the form $\partial^2 \phi^2 A^2$, but only two, 
\begin{align}
 \mathcal{L}_8 =  \frac{ \lambda_8 m^2}{\Lambda^4}  \left( A^\mu \partial_\mu \phi \right)^2 
+\frac{ \lambda_8' m^2}{\Lambda^4} \left( A_\mu \partial_\nu \phi \right)^2 
\label{eqn:VS_eg_L8}
\end{align}
are compatible with an approximate shift symmetry for $\phi$ (which is softly broken by its mass). Here we have explicitly included factors of the $A_\mu$ mass so that the interactions become strongly coupled at $\Lambda$ when $\lambda_8, \lambda_8' \sim \mathcal{O} (1)$ (i.e. the decoupling limit $m \to 0$ can be taken with $\Lambda, \lambda_8, \lambda_8'$ held fixed). 
The corresponding elastic amplitudes are,
\begin{align}
\Lambda^4 \mathcal{A}_{\pm \pm} (s,t) &=  \frac{-u \, m^2 }{s-4m^2} \left(  \lambda_8 t +2 \lambda_8' (  t- 2 m^2)    \right)   \; ,   \\
\Lambda^4 \mathcal{A}_{00} (s,t) &=  \frac{\lambda_8}{2} \frac{s}{s-4m^2} \left( u^2 + (s-4m^2)^2 \right)    
+ \lambda_8' ( t - 2 m^2  ) \left( 2m^2 + \frac{s t }{s-4m^2}   \right)    \; .  \nonumber 
\end{align}

Assuming the Froissart bound in the UV, there is one non-trivial positivity bound,
\begin{align}
 \partial_s^4 \left(   s ( s - 4 m^2 )  \mathcal{A}_{00} (s,t)   \right) |_{ \substack{ s=4m^2 \\ t = 0 } } = \frac{4! \lambda_8}{ \Lambda^4}  > 0 \;\;\;\; \left( \text{when} \;\; N_{\rm UV} = 1 \right) \, .
 \label{eqn:VS_eg_pos1}
\end{align}
The $\lambda_8'$ interaction is not bounded because it vanishes in the forward limit. 
In this simple example, what our new bound \eqref{eqn:pos_main2} shows is that, assuming a super-Froissart convergence \eqref{eqn:superFroissart2} of the UV amplitude, then $\lambda_8'$ must obey,
\begin{align}
 \partial_t  \left(   \frac{  s - 4 m^2 }{ - u }  \mathcal{A}_{\pm\pm} (s,t)   \right) \big|_{ \substack{ s=4m^2 \\ t = 0 } } = \frac{  \lambda_8 + 2 \lambda_8' }{ \Lambda^4}  > 0  \;\;\;\; \left( \text{when} \;\; N_{\rm UV} = 1/2 \right) \,  .
  \label{eqn:VS_eg_pos2}
\end{align}
For the transverse modes, $A^2 (\partial \phi )^2$ is effectively a dimension-6 operator, and here we are seeing that the sign of this operator is directly tied to the degree of convergence in the UV. 
We can make this all the more striking by considering the single operator,  $\tilde{\lambda}_8 A^\mu \partial_{[\mu} \phi A^\nu \partial_{\nu]} \phi$ (i.e. tuning $\lambda_8' = - \lambda_8$ in \eqref{eqn:VS_eg_L8}). Taken together,  \eqref{eqn:VS_eg_pos1} and \eqref{eqn:VS_eg_pos2} imply that $\tilde{\lambda}_8 > 0$ and $\tilde{\lambda}_8 < 0$, showing that this interaction can \emph{never} be generated in isolation by a unitarity, causal, local UV completion unless the corresponding UV amplitude grows at large momenta with $N_{\rm UV} > 1/2$ (e.g. contains the tree-level $t$-channel exchange of a vector field).

\subsection{Four-Fermion Interactions}

Finally, we consider a single Dirac spinor, $\psi$, with canonical kinetic and mass terms,
\begin{align}
\mathcal{L}_{\rm free} = - \frac{1}{2} \bar\psi \gamma^\mu \partial_\mu \psi - \frac{1}{2} m \bar{\psi} \psi \; . 
\label{eqn:FFeg_Lfree}
\end{align}
The leading interactions appear at dimension-6, and take the form $\bar{\psi} \Gamma_1 \psi \, \bar{\psi} \Gamma_2 \psi$, where $\Gamma_1$ and $\Gamma_2$ are products of $\gamma_\mu$ and $\gamma_5$ matrices---two such interactions are considered in Appendix~\ref{app:polarisation}.   
In this subection, we focus on theories with a fermionic shift symmetry, $\psi \to \psi +  \epsilon$, which is broken softly by the mass in \eqref{eqn:FFeg_Lfree}. This approximate symmetry suppresses all the interactions at dimension-6, and also those at dimension-8. The first interactions compatible with the symmetry are dimension-10 operators of the form $\left( \partial \bar{\psi} \partial \psi \right)^2$. 
Such shift-symmetric fermions arise in a variety of settings, including Goldstinos from broken supersymmetry, and the longitudinal mode of a massive spin-3/2 field in the decoupling limit (to which positivity bounds were recently applied in \cite{Melville:2019tdc}). 

While there are a handful of different possible contractions, we will focus on a single interaction to illustrate our key point,
\begin{align}
\mathcal{L}_{\rm int} = \lambda_{10} ( \partial_\mu \bar{\psi} \partial^\mu \psi )^2 \; .
\end{align}
Assuming a local UV completion, the forward-limit positivity bounds are\footnote{
The polarisation tensors can be found explicitly in Appendix~\ref{app:polarisation}, as well as a description of how the overall sign of the fermion amplitude is determined. 
},
\begin{align}
\begin{aligned}
&\partial_s^2 \hat{\mathcal{A}}_0^{\psi \psi \to \psi \psi} = - 4 m^2 \lambda_{10} > 0  \; ,  \\
&\partial_s^4 \hat{\mathcal{A}}^{\psi \bar{\psi} \to \psi \bar{\psi}}_{\pm} = + 48 m^2 \lambda_{10} > 0 \;  ,
\end{aligned}
 \;\;\;\; \left( \text{when  } N_{\rm UV} = 1 \right) 
\label{eqn:psi_eg_m}
\end{align}
which together require that $\lambda_{\rm 10} = 0$. 
It was already argued in~\cite{Bellazzini:2016xrt} that a fermion with exact shift-symmetry would inevitably violate the $\mathcal{A}^{(2)} > 0$ positivity bounds (analogous to how an exact Galileon symmetry violates this bound in scalar scattering). \eqref{eqn:psi_eg_m} shows that, even allowing for a small mass term to softly break the shift symmetry, this interaction still violates the positivity bounds required for a local UV completion. 

One possible response (which could also be made for the Galileon) is that whatever UV completes this shift-symmetric fermion must contain some degree of non-locality\footnote{
Another possible resolution is that shift-symmetric fermions never exist in isolation, but always come coupled to other light fields (which contribute to the $\psi \psi \to \psi \psi$ amplitudes to restore positivity). This parallels the case of a massless spin-3/2 field, whose amplitudes cannot be consistent unless coupled to a spin-2 in a supersymmetric way, i.e. the gravitino must always come with a graviton \cite{Grisaru:1977kk}.
}. 
Then the Froissart bound need not apply, and no positivity constraint can be placed on $\lambda_{10}$. One virtue of our improved positivity bound \eqref{eqn:pos_main2} is that, even allowing for some non-locality, in particular $N_{\rm UV} = 3/2$ and the sub-Froissart condition~\ref{eqn:subFroissart2}, there is still a constraint on this dimension-10 operator, 
\begin{align}
\partial_t \partial_s^2 \hat{\mathcal{A}}_0^{\psi \psi \to \psi \psi} = - \lambda_{10} > 0 \; \;\; \left( \text{when } N_{\rm UV} = 3/2 \right) \; . 
\end{align}
This final application of our stronger positivity bound is complementary to our earlier examples, in which \emph{stronger} UV convergence is used to constrain \emph{lower} dimension operators. 
By removing the need for the forward-limit $\mathcal{A}^{(2N)}|_{t=0}$ in our positivity bound for $\partial_t \mathcal{A}^{(2N)} |_{t=0}$, it is now possible to constrain a different range of interactions than previously possible (for a given UV behaviour of the amplitude).

\section{Discussion}
\label{sec:discussion}

It is now well-established that by assuming four fundamental properties of short-distance physics -- namely Lorentz invariance (LI), unitarity, causality, and locality -- one can place various bounds on scattering amplitudes at low energies.
Roughly, these properties have the following consequences for a $2\to 2$ scattering amplitude $\mathcal{A}$:
\begin{itemize}
	\item LI $\implies$ $\mathcal{A}$ is a function of the Mandelstam variables $s$ and $t$.
	\item Unitarity $\implies$ an optical theorem relating $\mathrm{Im}\,\mathcal{A}$ to a (positive) cross-section.
	\item Causality (+ LI) $\implies$ a domain of analyticity for $\mathcal{A}(s)$ in the complex $s$-plane.
	\item Locality (+ LI + unitarity) $\implies$ $\mathrm{lim}_{|s|\to \infty}|\mathcal{A}(s)|$ is polynomially bounded.
\end{itemize}
By recalling these standard ingredients for cooking up positivity bounds, we learn that the fundamental UV properties are most powerful when used {\em in combination} (as is best exemplified by the boundedness condition $\mathrm{lim}_{s\to \infty}\mathcal{A}(s) < s^2$ of Froissart--Martin).

In this spirit, we began the present work by deriving stronger unitarity conditions that `knead in' LI more thoroughly than in previous literature, and these stronger unitarity conditions led us to new positivity bounds. Specifically, for $2\to 2$ scattering of particles with spin, invariance of the theory under the rotation subgroup $SO(3)\subset SO(3,1)$ supplies us with {\em conserved angular momentum} quantum numbers that lead to selection rules in both the $s$- and $u$-channels. Firstly, by performing an $s$-channel partial wave expansion, one can factor out the kinematic part of the scattering amplitude for each value of the total angular momentum $J$, and write it as a matrix element of a rotation operator $\hat{J}_y$ (that rotates the axis of the incoming pair onto the axis of the outgoing pair). Moreover, when scattering incoming particles with helicities $h_1$ and $h_2$, this sum over $J$ begins at a finite value $J_\mathrm{min} = |h_1-h_2|$. 
By combining these simple facts with the optical theorem, we derived an ``angular momentum enriched'' unitarity condition, which is that, schematically\footnote{
For elastic helicity configurations, $\mathcal{A}_{h_1 h_2 h_1 h_2}$, the absorptive part defined in \eqref{eqn:Disc_s} coincides with the usual imaginary part in the physical $s$ region. 
} (see (\ref{eqn:unit_dtAs})), 
$$\partial_t \mathrm{Im} \,\mathcal{A} |_{t=0}> \frac{|h_1-h_2|}{s-4m^2} \mathrm{Im} \,\mathcal{A} |_{t=0}$$
in the physical $s$-channel region,
to be contrasted with the more familiar unitarity conditions, $\mathrm{Im}\,\mathcal{A} |_{t=0} >0$ and $\partial_t \mathrm{Im}\,\mathcal{A} |_{t=0}>0$.

To use such an improved unitarity condition to derive an EFT positivity bound, one also needs to derive a similar condition for the $u$-channel branch cut of $\mathcal{A}$. To do this one has to work a little harder, thanks to the complicated nature of the crossing relation when considering massive spinning particles. 
The obstacles to overcome here are for the most part technical, related to sign-indefinite terms that arise from taking a $t$-derivative of the matrices that implement crossing; only by averaging over all ingoing helicities (that have fixed $h_1+h_2$) do we obtain a positive quantity.
The $u$-channel unitarity condition is then, schematically (see (\ref{eqn:unit_dtAu})),
$$\partial_t \mathrm{Im}\,\sum_{\mathrm{hels.}}\mathcal{A} |_{t=0}> \frac{|h_1+h_2|}{u-4m^2} \mathrm{Im} \,\sum_{\mathrm{hels.}}\mathcal{A} |_{t=0} $$
in the physical $u$-channel region.

Armed with these new unitarity conditions, which are stronger than those used previously for massive particles with spin, we exploited analyticity of the $S$-matrix and polynomial boundedness of the UV amplitude to derive new positivity bounds involving one $t$-derivative and $2N$ $s$-derivatives of a $2\to 2$ elastic amplitude $\sum_{\mathrm{hels.}}\mathcal{A}$, summed over helicities as above. The bounds are most powerful for large helicities, because in that case the $s$- and $u$-channel branch cut contributions have the same sign. It thus helps to divide the analysis into two cases depending on whether the number
$$\alpha := 2N+1-|h_1+h_2| - |h_1-h_2|_\text{min}$$
is positive (small helicities) or negative (large helicities).
For $\alpha \leq 0$, the new bound is that, roughly (see (\ref{eqn:pos_main2})),
$$ \partial_t \partial_s^{2N} \sum_{\mathrm{hels.}}\mathcal{A}(s,t)|_{s=4m^2,\, t=0} > 0, $$
which holds for UV completions in which the amplitude is bounded by $\mathrm{lim}_{s\to \infty}\mathcal{A} < s^{2N+1}$. We also derive a new bound (\ref{eqn:pos_main1}) in the small helicity regime $\alpha >0$ (which is stronger than similar bounds previously derived in~\cite{deRham:2017zjm}), that is applicable for UV theories with faster convergence in the UV, namely $\mathrm{lim}_{s\to \infty}\mathcal{A} < s^{2N}$.

These very general new results hold for massive spinning particles with any spins and helicities. In the special case of (i) scattering spin-$1/2$ particles with opposite helicities, (ii) assuming the amplitude has a super-Froissart convergence $\mathrm{lim}_{s\to \infty}\mathcal{A} < s$, and (iii) in the limit that the spinors are massless, this reproduces the bound recently derived in \cite{Remmen:2020uze}. Indeed for general spins, our bounds simplify in the limit of massless particles; if there are no matrix elements that diverge as $m\to 0$,
then one obtains positivity of $ \partial_t \partial_s^{2N}\mathcal{A}$ without the need to sum over helicity combinations. We discuss a small selection of applications of the new bounds in \S \ref{sec:examples}. We save further investigation of the phenomenological implications of such bounds for future work.

Finally, we do not claim that the new bounds (\ref{eqn:pos_main1}) and (\ref{eqn:pos_main2}) are in any way the {\em strongest} positivity bounds that apply for massive spinning particles. Rather, we claim simply that the bounds are {\em stronger} than those proposed in previous literature, thanks to the angular momentum enriched unitarity conditions (\ref{eqn:unit_dtAs}) and (\ref{eqn:unit_dtAu}). 
In fact, in light of the recent progress made for scalar amplitudes, in particular the identification of optimal positivity bounds~\cite{Bellazzini:2020cot, Arkani-Hamed:2020blm, Chiang:2021ziz} and fully exploiting crossing symmetry~\cite{Tolley:2020gtv, Caron-Huot:2020cmc,  Caron-Huot:2021rmr}, it  seems likely that one may derive even stronger bounds on massive spinning amplitudes by combining these advances with the improved unitarity conditions and crossing relation given here. 
We close by briefly outlining how this could be achieved.

\subsection{Future Directions}
\label{sec:future}

\paragraph{Upper unitarity bound:}
In the small helicity regime, where $\alpha > 0$, the bound~\eqref{eqn:pos_main1} presented here represents a small numerical improvement over the analogous bounds in \cite{deRham:2017avq} and \cite{deRham:2017zjm}, but still involve the forward limit amplitude with zero $t$-derivatives. However, there is a way to further exploit unitarity to replace this contribution by a constant. 
Explicitly, inserting a complete set of angular momentum states in the unitarity condition~\eqref{eqn:unit_T} leads to the analogue of \eqref{eqn:optical} for the partial wave coefficients,
\begin{align}
2 \, \text{Im} \, a^{J_s}_{h_1 h_2} \geq \frac{1}{V_s} | a^{J_s}_{h_1 h_2} |^2 \;  ,
\label{eqn:unit_aJ}
\end{align}
where $a^{J_s}_{h_1 h_2} (s) := \langle h_1 h_2 |  \hat{T} (p_s, J_s , h_s )  | h_1 h_2 \rangle$ are the elastic coefficients appearing in \eqref{eqn:Apw}, and the factor of $1/V_s$ comes from resolving the identity using states normalised as in \eqref{eqn:PJM_norm}. 
Unitarity therefore requires not only that each $\text{Abs}_s \, a^{J_s}_{h_1 h_2}$ is positive, but also bounded from above\footnote{
Note that it is common to rescale the partial wave coefficients (i.e. the state normalisation \eqref{eqn:PJM_norm}), defining $f^J_{h_1 h_2} := \left[ 16 \pi (2J+1) \sqrt{s}/k_s \right]^{-1}  a^J_{h_1 h_2}$ so that \eqref{eqn:unit_aJ} is simply $\text{Abs}_s \, f^J_{h_1 h_2} \leq g_s$. We have instead chosen to keep the partial wave expansion simple (i.e. $|c_{JM}| = 1$ in \eqref{eqn:Apw}) which leads to an explicit $V_s$ phase space factor in the unitarity bound.
},
\begin{align}
2 V_s \geq | a^{J_s}_{h_1 h_2} (s) |  \geq  \text{Im} \, a^{J_s}_{h_1 h_2} (s)   \geq 0 \; ,
\label{eqn:unit_aJs}
\end{align}
and similarly in the $u$-channel.  
It is the upper bound in \eqref{eqn:unit_aJs} 
which can be used to place further positivity bounds on the $t$-derivative of the EFT amplitude, particularly in the small helicity regime. 
For instance, using the partial wave expansion in the dispersion relation~\eqref{eqn:dt_disp},
\begin{align}
2 u \, I_u \geq \sum_{J_u = |h_u|}^{\infty} \big[ 
J_u (J_u + 1) - |h_u| \left( |h_u| + 1 \right) - \alpha    
 \big]  \sum_{ \substack{ h_1 , \, \bar{h}_2 \\ h_1 - \bar{h}_2 = h_u } }  \text{Im} \, a^{J_u}_{h_1 \bar{h}_2}  \; ,
\end{align}
we see that even when $\alpha > 0$ it is only the first few partial waves which spoil positivity of the $u$-channel branch cut. Defining $J_* \geq |h_u|$ as the smallest half-integer for which the quantity $J_u (J_u + 1) - |h_u| \left( |h_u| + 1 \right) - \alpha$ is positive, we can split the sum into $\sum_{J_s = |h_u|}^{J_*-1}$ and $\sum_{J_s=  J_*}^{\infty}$, and then use $\text{Im} \, a^{J_u}_{h_1 \bar{h}_4} \geq 0$ for the latter and $\text{Im} \, a^{J_u}_{h_1 \bar{h}_4} \leq 2 V_u$ for the former. This produces a bound of the form,
\begin{align}
\partial_t \partial_s^{2N} \sum_{\rm hels.} \mathcal{A} (s,t) |_{\substack{s=4m^2 \\ t=0}} >  \text{negative constant} \; ,
\label{eqn:const_bound}
\end{align}
which holds for any value of $\alpha$. 
For instance, for scattering two distinguishable scalars, $\phi \varphi \to \phi \varphi$, the dispersion relation for the third derivative $( \tfrac{1}{2} \partial_t \partial_s^2 - \tfrac{1}{3!} \partial_s^3 ) \mathcal{A}$ receives a negative contribution only from the $J_s = 0$ and $J_u = 0$ partial waves, and so,
\begin{align}
\left( \tfrac{1}{2} \partial_t \partial_s^2 - \tfrac{1}{3!} \partial_s^3 \right)  \mathcal{A} (s,t) |_{\substack{s=4m^2 \\ t=0}} &> 
- \tfrac{3}{2} \int_\mu^{\infty} \frac{ds}{\pi} \frac{16 \pi \sqrt{s} }{ (s-4m^2)^{9/2}}  
- \tfrac{3}{2} \int_\mu^{\infty} \frac{du}{\pi} \frac{16 \pi }{u^{7/2} \sqrt{u-4m^2}}  \nonumber \\
&= - \frac{16}{ \mu^3} + \mathcal{O} \left( \frac{4m^2}{\mu} \right) \; .
\label{eqn:const_bound_eg}
\end{align}
In perturbation theory, thanks to the trivial $s \leftrightarrow u$ crossing relation, the $\phi \varphi \to \phi \varphi$ scalar amplitude at tree-level is freely generated by $t$ and $s^2 + u^2$, 
\begin{align}
\mathcal{A} (s,t) = \sum_{a, \, b} \, g_{2a , b} \, (s^2 + u^2 )^a t^b \; ,
\label{eqn:Aexp_general}
\end{align}
and the bound \eqref{eqn:const_bound_eg} requires that the Wilson coefficient $g_{2,1} > - 8 /\mu^3$. 
This effectively places a limit on the cut-off of the EFT. 
For comparison, the largest partial wave coefficient from this term is $a^0_{00} (s) = - 7 g_{2,1} s^3/12$ in the EFT, and so perturbative unitarity requires that $|g_{2,1} s^3 | <  192 \pi/7 \approx 86$. 
Our bound \eqref{eqn:const_bound_eg}, which incorporates unitarity, analyticity and locality, thus improves numerically on this partial wave constraint by an order of magnitude. 
This strategy of removing a finite number of partial waves using the upper bound imposed by elastic unitarity can clearly be further optimised, and we leave that direction open for future exploration. 

\paragraph{Full crossing symmetry:}
For the scattering of identical scalars, the $t$-channel process $1 \bar{3} \to \bar{2} 4$ is also elastic and therefore has partial wave coefficients bounded by unitarity. 
Fully exploiting this additional crossing relation leads to so-called ``null constraints'' (for instance $g_{4,0} = g_{2,2}$ in \eqref{eqn:Aexp_general}) that can be used to improve the positivity bounds \cite{Tolley:2020gtv, Caron-Huot:2020cmc,  Caron-Huot:2021rmr}. 
Such null constraints have not yet been developed for massive spinning particles, where the crossing relation is more complicated and the null constraints must mix different helicity configurations. 
In particular, the $t$-channel image of $\mathcal{A}_{h_1 h_2 h_1 h_2}$ no longer has an elastic helicity configuration, even when the particles are massless (unless $h_1 = h_2 = 0$). 
While preparing this manuscript, a step in this direction was taken by \cite{Henriksson:2021ymi} for light-by-light scattering, and it would be interesting to extend this systematically to other spins and include the effects of a finite mass.

\paragraph{The moment problem:}
For the case of scalar amplitudes, there has been much progress in identifying a set of optimal positivity bounds using probability theory, using the dispersion relation to relate EFT derivatives to the moment problem~\cite{Bellazzini:2020cot, Arkani-Hamed:2020blm, Chiang:2021ziz}. 
The resulting bounds become particularly useful for higher-order $s$ and $t$ derivatives, which must obey a tower of Hankel determinant conditions. 
In Appendix~\ref{app:higher} we describe how the selection rule $J_s \geq |h_s|$ naturally leads to an infinite tower of improved unitarity bounds on every $t$-derivative of the $s$-channel branch cut, and in particular we identify the differential operators which correspond to the matrix elements $\langle J_s \; h_s | \hat{J}_y^{2n} | J_s \; h_s \rangle$.
We hope that these selection rules and improved unitarity conditions at higher orders will facilitate future connections with the moment problem for massive spinning particles.

\subsection*{Acknowledgments}
We thank the members of the Cambridge Pheno Working Group for useful discussions.
SM is supported by an UKRI Stephen Hawking Fellowship (EP/T017481/1), and 
TY is supported by a Branco Weiss Society in Science Fellowship.
This work was partially supported by the STFC consolidated grants ST/P000681/1 and ST/T000694/1.

\appendix
\section{Higher \boldmath$t$ Derivatives}
\label{app:higher}

In the main text, we have considered positivity bounds on the first $t$-derivative of the $2\to2$ amplitude for massive spinning particles. 
The key to strengthening these bounds beyond the forward limit lay in the selection rule \eqref{eqn:Jy2_identity} for the matrix element $\langle \hat{J}_y^2 \rangle$ and the resulting bound \eqref{eqn:unit_dtAs} on $\partial_t \text{Abs}_s \, \mathcal{A}_{h_1 h_2 h_1 h_2}$. 
In this Appendix, we discuss the generalisation of these identities to higher $t$-derivatives.  

Let us begin by defining the dimensionless variable,
\begin{align}
 \hat{t} := \frac{t}{s-4m^2}
\end{align}
so that $\partial_{\hat{t}} = (s-4m^2) \partial_t$.
As shown in section~\ref{sec:unitarity}, $\partial_{\hat{t}} \, \mathcal{A}_{h_1 h_2 h_3 h_4}$ is proportional to the matrix element, $\langle J_s \, h_s |  \hat{J}_y^2 | J_s \, h_s \rangle$. 
The key idea is that, with the partial wave expansion,
\begin{align}
\mathcal{A}_{h_1 h_2 h_3 h_4} (s,t) = \sum_{J_s} \langle J_s \; h_s^{\rm out} |  e^{- i \hat{J}_y \theta_s} | J_s \; h_s^{\rm in} \rangle \; a^{J_s}_{h_1 h_2 h_3 h_4} (s) 
\end{align}
where $a^{J_s}_{h_1 h_2 h_3 h_4} (s) := \langle h_3 h_4 | \hat{T} (s, J_s, h_s^{\rm in} )  | h_1 h_2\rangle$ are the partial wave coefficients, higher $\theta_s$ derivatives can be used to produce all $\langle \hat{J}_y^{2n}\rangle$ matrix elements,
\begin{align}
(-1)^n \partial_{\theta_s}^{2n} \mathcal{A}_{h_1 h_2 h_3 h_4} (s,t) |_{t=0} = \sum_{J_s} \langle J_s \; h_s^{\rm out} |  \hat{J}_y^{2n} | J_s \; h_s^{\rm in} \rangle \; a^{J_s}_{h_1 h_2 h_3 h_4} (s)  \; ,
\end{align}
where in terms of $t$, 
\begin{align}
\partial_{\theta_s}^{2n} = \left( - \sqrt{t u} \, \partial_t \right)^{2n} = \sum_{j=1}^n c_{n,j} \partial_{\hat t}^j \; \text{ at } t=0 \; , 
\end{align}
where $c_{n,j}$ are fixed constants. 
For instance, for $n=3$, 
\begin{align}
\left( \frac{15}{8} \partial_{\hat{t}}^3 
 + \frac{15}{4} \partial_{\hat{t}}^2 
 + \frac{1}{2} \partial_{\hat{t}} \right)
 \mathcal{A}_{h_1 h_2 h_3 h_4} (s,t) |_{t=0} = \sum_{J_s } \langle J_s h_s^{\rm out} |  \hat{J}_y^{6} | J_s h_s^{\rm in} \rangle \; a^{J_s}_{h_1 h_2 h_3 h_4} (s) \, .
\end{align}
Conceptually, these $\langle \hat{J}_y^{2n} \rangle$ matrix elements satisfy inequalities analogous to \eqref{eqn:Jy2_identity} as a result of the selection rule $J_s \geq |h_s|$, and these inequalities can be used to place bounds between the different $\partial_{\theta_s}^{2n} \text{Abs}_s \mathcal{A}_{h_1 h_2 h_1 h_2}$. 

In particular, the $\langle J \; h | \hat{J}_y^{2n} | J \; h \rangle$ have the property that they grow monotonically with $J$, with a minimum value at $J=h$. For example,
\begin{align}
\langle J \; h | \hat{J}_y^2 | J \; h \rangle &= \tfrac{1}{2} \left( \mathcal{J} - h^2 \right)   = \tfrac{1}{2} h  \text{ at } J=h  \; ,  \nonumber \\
\langle J \; h | \hat{J}_y^{4} | J \; h \rangle &= \tfrac{1}{8} \left(
3 \mathcal{J}^2
- 2  \mathcal{J} ( 1 + 3 h^2)
+ h^2 (5 + 3 h^2)  \right)
  = \tfrac{3}{4} h^2 - \tfrac{1}{4} h  \text{ at } J=h \; , \nonumber \\
\langle J \; h | \hat{J}_y^{6} | J \; h \rangle &= \tfrac{1}{16} \left(
5 \mathcal{J}^3
- 5 \mathcal{J}^2 ( 2 + 3 h^2 )
+ \mathcal{J} ( 8 + 45 h^2 + 15 h^4 )
- h^2 \left( 28  + 35 h^2 + 5 h^4 \right)  \right)
  \nonumber \\
 &=  \tfrac{15}{8}  h^3  - \tfrac{15}{8}  h^2 +  \tfrac{1}{2} h \text{ at } J=h \; ,  
\end{align} 
where $\mathcal{J} = J(J+1)$. 
It is straightforward to construct linear combinations of the $\hat{J}_y^{2n}$ such that these minimum values are simply $h^n$, i.e.,
\begin{align}
\hat{\mathcal{O}}_n := \sum_j C_{n,j} \hat{J}_y^{2n}  
\end{align}
with coefficients $C_{n,j}$ chosen so that $\langle J \; h | \hat{\mathcal{O}}_n | J \; h \rangle = h^n$ at $J=h$.  These combinations also grow monotonically with $J$, and appear to satisfy the analogue of \eqref{eqn:Jy2_identity}, 
\begin{align}
\langle J \; h | \hat{\mathcal{O}}_n | J \; h \rangle \geq |h| \, \langle  J \; h | \hat{\mathcal{O}}_{n-1} | J \; h \rangle \; ,
\label{eqn:On_identity} 
\end{align}
which we have checked numerically for all $2n$ up to $10$. 
For instance, the first few are,
\begin{align}
\hat{\mathcal{O}}_0 = 1 \; , \;\;  \hat{\mathcal{O}}_1 = 2 \hat{J}_y^2 \; , \;\; \hat{\mathcal{O}}_2 = \frac{4}{3} \hat{J}_y^2 + \frac{2}{3} \hat{J}_y^2 \; , \;\; \hat{\mathcal{O}}_3 = \frac{8}{15} \hat{J}_y^6 + \frac{4}{3} \hat{J}_y^4 + \frac{2}{15} \hat{J}_y^2  \; . 
\end{align}
In order to translate \eqref{eqn:On_identity} into bounds on $\text{Abs}_s \, \mathcal{A}_{h_1 h_2 h_1h_2}$, we define $D_t^n$ as the linear combination of $\hat{t}$-derivatives which produces $\hat{\mathcal{O}}_n$, 
\begin{align}
D_t^n \mathcal{A}_{h_1 h_2 h_3 h_4} (s,t) |_{t=0} = \sum_{J_s} \langle J_s \, h_s^{\rm out} |  \hat{\mathcal{O}}_n | J_s \, h_s^{\rm in} \rangle \; a^{J_s}_{h_1 h_2 h_3 h_4} (s)  \; ,
\label{eqn:DtnA}
\end{align}
so that,
\begin{align}
D_t^n \text{Abs}_s \, \mathcal{A}_{h_1 h_2 h_1h_2} (s,t) |_{t=0} \geq |h_s| \; D_t^{n-1} \text{Abs}_s \,\mathcal{A}_{h_1 h_2 h_1h_2} (s,t) |_{t=0} \; , 
\label{eqn:unit_As_general}
\end{align}
for any physical value of $s>4m^2$, since $\text{Abs}_s a^{J_s}_{h_1 h_2 h_1 h_2} (s) > 0$. 
\eqref{eqn:unit_As_general} is the extension of the $s$-channel unitarity bound \eqref{eqn:unit_As} to arbitrary $t$-derivatives. 

The $D_t^n$ derivative operators in \eqref{eqn:DtnA} can be written explicitly,
\begin{align}
 D_t^n = \sum_{k=0}^n  \left\{ \begin{array}{c} n \\ k \end{array} \right\}  \partial_{\hat{t}}^k
 \label{eqn:DtnDef}
\end{align}
where $\scriptsize \left\{ \begin{array}{c} n \\ k \end{array} \right\} $ are the Stirling triangle numbers of the second kind, which are the coefficients in the expansion,
\begin{align}
 x^n = \sum_{k=0}^n \left\{ \begin{array}{c} n \\ k \end{array} \right\}  (x)_k \;\;\;\; \text{where} \;\; (x)_k = x (x-1) ... ( x- k +1 ) \; ,
\end{align}
and are related to the usual binomial coefficients by,
\begin{align}
\left\{ \begin{array}{c} n \\ k \end{array} \right\}  = \frac{1}{k!} \sum_{i=0}^k (-1)^i 
\left( \begin{array}{c} k \\ i \end{array} \right) (k-i)^n \; .
\end{align}
The first few inequalities beyond \eqref{eqn:unit_As} are therefore,
\begin{align}
\text{Abs}_s \, \mathcal{A}_{h_1 h_2 h_1 h_2} (s,t) |_{t=0} 
&\leq   |h_s|^{-1}\left(  \partial_{\hat{t}} \right)  \text{Abs}_s \, \mathcal{A}_{h_1 h_2 h_1 h_2} (s,t) |_{t=0}  \nonumber \\ 
&\leq |h_s|^{-2} \left( \partial_{\hat{t}}^2 +  \partial_{\hat{t}}  \right) \text{Abs}_s \, \mathcal{A}_{h_1 h_2 h_1 h_2} (s,t) |_{t=0}  \nonumber \\
&\leq |h_s|^{-3}  \left( \partial_{\hat{t}}^3 + 3 \partial_{\hat{t}}^2 +  \partial_{\hat{t}} \right)  \text{Abs}_s \, \mathcal{A}_{h_1 h_2 h_1 h_2} (s,t) |_{t=0}  \nonumber \\ 
&\leq |h_s|^{-4}  \left( \partial_{\hat{t}}^4 + 7 \partial_{\hat{t}}^3 + 6 \partial_{\hat{t}}^2 + \partial_{\hat{t}} \right) \text{Abs}_s \, \mathcal{A}_{h_1 h_2 h_1 h_2} (s,t) |_{t=0} \;, 
\end{align}
and so on, to arbitrary orders in $t$ derivatives. 

However, crossing the identities \eqref{eqn:unit_As_general} to the $u$-channel is quite involved since the higher order $t$-derivatives also act on the crossing matrices in \eqref{eqn:Across_pw} and produce sign-indefinite terms analogous to \eqref{eqn:Sy2}, which must be carefully combined into something positive. In this work we have focussed on the first ($n=1$) identity and have shown that the helicity sum \eqref{eqn:AhuDef} guarantees the $u$-channel bound \eqref{eqn:unit_dtAu}, leading to stronger positivity bounds on the EFT dispersion relation. 
A systematic study of how to do this for the $n > 1$ identities is left open for the future. 

\paragraph{Connection with the moment problem:}
In addition to \eqref{eqn:unit_As_general}, there are further inequalities which must be satisfied by these combinations of $t$ derivatives in order to solve the corresponding moment problem. 
A similar observation was made recently in \cite{Chiang:2021ziz} for scalar amplitudes.
In our case, the regulated amplitude for massive spinning particles defined in \eqref{eqn:Ahath1h2} has $t$-derivatives given by,
\begin{align}
\frac{1}{k!} \partial_{\hat{t}}^k \, \hat{\mathcal{A}}_{h_1 h_2 h_1 h_2} (s,t) |_{t=0} = \sum_{J_s = |h_s|}^{\infty} v_{J_s , |h_s| , k} \;   a^{J_s}_{h_1 h_2 h_1 h_2} (s)
\end{align}
where the coefficients $v_{J, h, k}$ are,
\begin{align}
 v_{J , h,  k} 
 := \frac{ \prod_{a=h+1}^{h+k} \left( \mathcal{J} - a (a-1) \right)  }{(k!)^2}  \; ,
\end{align}
and reduce to the well-known expression for the derivatives of the Legendre polynomials $P_{J}^{(k)} (1)$ when $h=0$.
A linear combination of $\hat{t}$-derivatives of $\hat{A}_{h_1 h_2 h_3 h_4}$ can therefore be constructed to give a partial wave expansion of simply $\mathcal{J}^n_s a^{J_s}_{h_1 h_2 h_1 h_2}$.  
Rather than $\mathcal{J}_s = J_s ( J_s + 1)$, which begins at $|h_s| (|h_s| + 1)$, it is more convenient to consider moments of $\mathcal{L}_s = J_s (J_s + 1) - |h_s| (|h_s| + 1 )$, which takes discrete values $\geq 0$ that are linear in $|h_s|$, namely $\mathcal{L}_s = 0$, $2 + 2 |h_s|$, $6 + 4 |h_s|$, $12 + 6 |h_s|$, $20 + 8 |h_s|$, $... $. 
Denoting by $\mathcal{D}_{\hat{t}}^n$ the combination of $\hat{t}$-derivatives that achieves, 
\begin{align}
 \mathcal{D}_{\hat{t}}^n \, \hat{\mathcal{A}}_{h_1 h_2 h_1 h_2} (s,t) |_{t=0} = \sum_{J_s = |h_s|}^{\infty}  \mathcal{L}^n_s \;   a^{J_s}_{h_1 h_2 h_1 h_2} (s) \; ,
\end{align}
the first few can be written explicitly as\footnote{
An explicit expression for the $\mathcal{D}_{\hat{t}}^n$ can be written analogously to \eqref{eqn:DtnDef} using suitably generalised Stirling numbers.
},
\begin{align}
\left(  \begin{array}{c} 
\mathcal{D}_{\hat{t}}^1  \\
\mathcal{D}_{\hat{t}}^2  \\
\mathcal{D}_{\hat{t}}^3  \\
\vdots
\end{array} \right) =
\begingroup\BigColSep
 \left( \begin{array}{c c c c}
 1 & 0 & 0 & \cdots \\
 2 + 2|h_s|   & 4 & 0 & \cdots \\
 4  (1 + |h_s| )^2  & 32 + 24 |h_s| & 36 & \cdots  \\
\vdots & \vdots & \vdots & \ddots 
 \end{array}   \right) 
\endgroup
 \left(  \begin{array}{c}
\partial_{\hat{t}}  \\
\tfrac{1}{2} \partial_{\hat{t}} \\
\tfrac{1}{3!} \partial_{\hat{t}} \\
\vdots
\end{array}  \right)
\end{align}
and coincide with the GL transformation given in \cite{Chiang:2021ziz} when $|h_s| =0$. 

We believe that identifying these particular combinations of $t$-derivatives, which isolate either a particular $\langle \hat{J}_y^{2n}\rangle$ or a particular $\mathcal{L}_s^n$ in the partial wave expansion, is the first steps towards developing an optimal set of positivity bounds for massive spinning particles using the moment theorems recently introduced in~\cite{Bellazzini:2020cot} for scalar amplitudes.
Further exploration in this direction is postponed for the future.

\section{Crossing Relation Details}
\label{app:crossing}

The crossing relation for massive spinning particles is given in \cite{Trueman:1964zzb, cohen-tannoudji_kinematical_1968, Hara:1970gc, Hara:1971kj} (see also \cite{deRham:2017zjm}), 
\begin{equation}
\mathcal{A}^{\psi_1 \psi_2 \to \psi_3 \psi_4}_{h_1 h_2 h_3 h_4} ( s, t )
=   \sum_{h_a'} \;  {\bf C}_{h_1 h_2 h_3 h_4}^{h'_1 h'_2 h'_3 h'_4} (\chi_u)  \;\;  \mathcal{A}^{ \psi_1 \bar{\psi}_4 \to \psi_3 \bar{\psi}_2 }_{h_1' h_4'h_3' h_2'} (u, t )     ,
 \label{eqn:helicityCrossing}
\end{equation}
where the crossing matrix can be decomposed into rotations of each of the four particles,
\begin{equation}
{\bf C}_{h}^{h'} (\chi_u)   =  \eta_u \, (-1)^{2S_2}\,  e^{i \pi( h_1'- h_3')}   d^{S_1}_{h_1' h_1} ( \chi_u )  d^{S_2}_{ h_2' h_2} ( -\pi + \chi_u )   d^{S_3}_{ h_3' h_3} (- \chi_u )  d^{S_4}_{ h_4' h_4} ( \pi -  \chi_u )
\label{eqn:Cmatrix}
\end{equation}
where the crossing angle $\chi_u$ is given in \eqref{eqn:chiu}, and $\eta_u$ is an overall sign which depends on the spin-statistics of the four particles\footnote{
In \cite{deRham:2017zjm}, it was shown that $\eta_u = (-1)^{S_1 + S_2 - S_3 + S_4} \eta_{12} \eta_{14} \,  \eta_{32} \eta_{34} \, \eta_{24}$ where $\eta_{ij} = -1$ if both $\psi_i$ and $\psi_j$ are fermions (so anticommute) and $=+1$ otherwise. Note that this sign depends on the choice of branch cuts for the Wigner $d$-matrices (which are only periodic in $4\pi$ for fermions), as pointed out in \cite{Hara:1970gc, Hara:1971kj}. We adopt the same convention as \cite{deRham:2017zjm}, in which the angles were chosen so that $0 \leq \theta_s < \pi$, $0 \leq \theta_u < \pi$ and $- \pi/2 \leq \chi_u \leq 0$. 
}. 
For the elastic processes we will consider (i.e. $S_1 = S_3$ and $S_2 = S_4$), this factor is $\eta_u = +1$.  
Note that the overall $(-1)^{2S_2}$ sign then simply encodes the statistics of particles 2 and 4, since this crossing can be thought of as the three permutations: (a) $2 \leftrightarrow 3$, (b) $2 \leftrightarrow 4$ and then (c) $4 \leftrightarrow 3$. The signs introduced by (a) and (c) cancel, leaving just $(-1)^{2S_2}$ introduced by (b).  

Rather than reproduce the rigorous proofs of this relation (which are somewhat involved), we simply sketch the three key steps:

\begin{itemize}

\item[(i)] \emph{Permute fields / CPT relation.}
The amplitude $\mathcal{A}_{h_1 h_2 h_3 h_4} (s,t)$ is related to the (LSZ reduction of the) time-ordered correlator $\langle T \, \hat{\psi}_{h_1} ( p_1) \hat{\psi}_{h_2} ( p_2 ) \hat{\psi}_{h_3} (p_3) \hat{\psi}_{h_4}  ( p_4) \rangle$. 
The crossing of particle $2$ from the in-state and $4$ from the out-state first requires permuting the fields $\hat{\psi}_{h_2} ( p_2)$ and $\hat{\psi}_{h_4} ( p_4)$ and using the CPT relation, which produces the spin-statistics factor $\eta_u$. 
 
\item[(ii)] \emph{Complex Lorentz transformation.}
While a straightforward exchange of $\hat{\phi}_{h_2} (p_2)$ and $\hat{\phi}_{h_4} (p_4)$ does formally exchange $s = - (p_1 + p_2)^2$ with $u = - (p_1 + p_4)^2$, it produces unphysical kinematics---in particular when $s$ is analytically continued to negative values, each $\bfp^s_a$ is in fact complex. 
To remedy this, we perform a (complex) Lorentz transformation to the frame \eqref{eqn:kinematics_u}. 
The explicit form of this transformation is given in \cite{Trueman:1964zzb,cohen-tannoudji_kinematical_1968,Hara:1970gc}. In particular, its action on each particle can be written as the product of a rest-frame rotation through a real angle and a complex boost, so that in the $u$-channel region,  
\begin{align}
 \mathcal{A}_{h_1 h_2 h_3 h_4} (s,t) &=  
\eta_u
\langle \bfp_3^u h_3 | e^{- i \hat{S}_{y}^{(3)} \chi_3} \langle \bfp_2^u \bar{h}_2 | e^{- i \hat{S}_{y}^{(2)} \chi_2 } \; \hat{T}  \; e^{i \hat{S}_{y}^{(1)} \chi_1 }  | \bfp_1^u h_1 \rangle e^{i \hat{S}_{y}^{(4)} \chi_4} | \bfp_4^u \bar{h}_4 \rangle  
\label{eqn:restFrameRotation}
\end{align}
where each $\hat{S}_{y}^{(a)}$ is a rotation in the plane of the scattering that acts only on each particle in its rest frame. For identical particle masses, these angles are given by,
\begin{align}
\label{eqn:AngleChoice}
\chi_1 = \chi_4 = - \chi_3 = -\chi_2  = - \chi_u  \,,
\end{align}
in terms of the $\chi_u$ defined in \eqref{eqn:chiu}. 
We cannot directly apply the partial wave expansion \eqref{eqn:state_pw} since now the 2-particle state $e^{i \hat{S}_{y} \chi_1 }  | \bfp_1^u h_1 \rangle e^{i \hat{S}_{y} \chi_4} | \bfp_4^u \bar{h}_4 \rangle  $ is no longer an eigenstate of global $\hat{J}_z$ rotations. However, we can insert a complete set of helicity states on either side of $\hat{T}$ like so, 
\begin{align}
 \mathcal{A}_{h_1 h_2 h_3 h_4} (s,t) &=  
\eta_u
\sum_{h'_a} \bar{C}_{h_3 \bar{h_2}}^{h_3' \bar{h}_2'} \langle \bfp_3^u h_3' |  \langle \bfp_2^u \bar{h}_2' |  \; \hat{T}  \; | \bfp_1^u h_1' \rangle | \bfp_4^u \bar{h}_4' \rangle 
C_{h_1 \bar{h_4}}^{h_1' \bar{h}_4'} \; . 
\end{align}

\item[(iii)] \emph{Reverse scattering plane normal.}
We have almost arrived at the $u$-channel kinematics, however now the normal to the scattering plane (defined for instance using $n_\mu = \epsilon_{\mu\alpha \beta\gamma} p_1^\alpha p_2^\beta p_3^\gamma$) has the opposite direction, i.e. $\hat{y} \to - \hat{y}$. This is the reason that the partial wave expansion \eqref{eqn:Across_pw} has the opposite sign of $\hat{J}_y$ to \eqref{eqn:Apw}. 

\end{itemize}

These steps lead to the expression \eqref{eqn:Across_pw} for the amplitude in the $u$-channel region, which can also be written as,
\begin{align}
\mathcal{A}^{\psi_1 \psi_2 \to \psi_3 \psi_4}_{h_1 h_2 h_3 h_4} ( s, t )
= \eta_u \sum_{h_a'} \, & d^{S_1}_{h_1' h_1} ( \chi_u) d^{S_2}_{\bar{h}_4' \bar{h}_4 } ( \chi_u)  
  d^{S_1}_{h_3 h_3'} ( \chi_u) d^{S_2}_{\bar{h}_2 \bar{h}_2' } ( \chi_u)    \nonumber \\
& \times (-1)^{ h_1' - \bar{h}_4' - h_3' + \bar{h}_2' } \mathcal{A}^{ \psi_1 \bar{\psi}_4 \to \psi_3 \bar{\psi}_2 }_{h_1' \bar{h}_4' h_3' \bar{h}_2'} (u, t )   \;  ,
\end{align}
where we have used the identity,
\begin{align}
d^{J_u}_{h_u^{\rm out} h_u^{\rm in}} ( - \theta_u ) = (-1)^{  h_u^{\rm in} - h_u^{\rm out}  } d^{J_u}_{h_u^{\rm out} h_u^{\rm in}} ( + \theta_u ) \; , 
\end{align}
in \eqref{eqn:Au_pw} to replace the partial wave sum over $J_u$ with $ \mathcal{A}^{ \psi_1 \bar{\psi}_4 \to \psi_3 \bar{\psi}_2 }_{h_1' \bar{h}_4' h_3' \bar{h}_2'} (u, t )$. 
This is equivalent to the expression \eqref{eqn:helicityCrossing} which appears throughout the literature, which can be easily verified using properties of the Wigner $d$ matrices\footnote{
See for instance the Appendix of \cite{deRham:2017zjm} for a list of useful Wigner $d$ matrix relations. 
}, 
\begin{align}
 d^{S_4}_{\bar{h}_4' \bar{h}_4} (  \chi_u) &= ( -1)^{S_4 - h_4'} d^{S_4}_{\bar{h}_4' h_4} ( \pi - \chi_u)   \; , \nonumber \\
 d^{S_3}_{ h_3 h_3' } (  \chi_u) &=  d^{S_3}_{h_3' h_3} ( - \chi_u)  \; ,  \nonumber \\
 d^{S_2}_{ \bar{h}_2 \bar{h}_2'} (  \chi_u) &= (-1)^{S_2 + h_2'} d^{S_2}_{\bar{h}_2' h_2} ( -\pi + \chi_u )    \; . 
\end{align}

Note that in the forward limit $t \to 0$ (i.e. $\chi_u \to 0$), the crossing relation \eqref{eqn:helicityCrossing} becomes simply,
\begin{align}
\mathcal{A}^{\psi_1 \psi_2 \to \psi_3 \psi_4}_{h_1 h_2 h_3 h_4} (4m^2- u ,0) = \mathcal{A}^{\psi_1 \bar{\psi}_4 \to \psi_3 \bar{\psi}_2 }_{h_1 \bar{h}_4 h_3 \bar{h}_2 } (u, 0) \; .
\label{eqn:crossingForward}
\end{align}
This facilitates the derivation of positivity bounds on $\partial_s^{2N} \mathcal{A}_{h_1 h_2 h_1 h_2} |_{t=0}$, since the $u$-channel branch cut ($\text{Abs}_u \, \mathcal{A}_{h_1 \bar{h}_2 h_1 \bar{h}_2} (u,t)$) is immediately positive by unitarity. 

Away from the forward limit, applying the crossing relation~\eqref{eqn:helicityCrossing} to the $u$-channel branch cut leads to a sum over inelastic amplitudes which is generally not positive. 
In \cite{deRham:2017zjm}, the crossing relation was diagonalised by transforming from states of definite helicity $|S \; h \rangle$  (eigenstates of $p \cdot \hat{J}$, rotations about particle momentum) to definite transversity $|S \; \tau \rangle$ (eigenstates of $\hat{J}_y$, rotations about normal of scattering plane), 
\begin{align}
 | S \; \tau \rangle := \sum_{h} \, u^S_{\tau h} \; | S \; h \rangle \; , 
\end{align}
 where the unitary matrix $u^S_{\tau h} = D^S_{\tau h} ( \tfrac{\pi}{2}, \tfrac{\pi}{2}, - \tfrac{\pi}{2} )$ implements the required rotation of spin quantisation axis. This provides a simple representation of the angular momentum matrix elements appearing in the partial wave expansion and crossing relation, 
\begin{align}
\langle S_2 \; h_2 | e^{- i \hat{J}_y \theta} | S_1 \; h_1 \rangle = \sum_\tau   u^{S_2}_{h_2 \tau} \,  e^{- i \tau \theta} \,  u^{S_1 \; *}_{\tau h_1} \; , 
\end{align}
and consequently the crossing of $\mathcal{A}_{\tau_1 \tau_2 \tau_3 \tau_4}  = \sum_{h_a}  u^{S_3 \, *}_{h_3 \tau_3} u^{S_4 \, *}_{h_4 \tau_4} \mathcal{A}_{h_1 h_2 h_3 h_4} u^{S_1}_{\tau_1 h_1} u^{S_2}_{\tau_2 h_2}$ is particularly simple,
\begin{align}
 \mathcal{A}_{\tau_1 \tau_2 \tau_3 \tau_4}^{\psi_1 \psi_2 \to \psi_3 \psi_4} (s,\theta_s) = \eta_u'  \, \exp \left(  - i  \chi_u \sum_a \tau_a  \right) \,  \mathcal{A}_{\tau_1 \tau_4 \tau_3 \tau_2}^{\psi_1 \bar{\psi}_4 \to \psi_3 \bar{\psi}_2} (u, -\theta_u )
\end{align}
since the rest frame rotation matrices $C^{h_1' \bar{h}_4'}_{h_1 \bar{h}_4}$ become simply an overall phase. The statistics factor $\eta_u'$ can be found explicitly in \cite{deRham:2017zjm}.

Rather than use this transversity basis, in this work we have remained in the helicity basis throughout. 
This allows us to leverage the selection rules, $J_s \geq |h_s|$ and $J_u \geq |h_u |$. 
As we have shown in section~\ref{sec:crossing}, the first $t$-derivative of the helicity crossing relation is positive for the particular helicity sum~\eqref{eqn:AhuDef}.   
In order to leverage the higher $t$-derivative bounds from Appendix~\ref{app:higher} to produce stronger positivity bounds on higher-order Wilson coefficients, one must identify an appropriate combination of helicity amplitudes for which the higher-order $\langle \hat{S}_y^{2n} \rangle$ matrix elements (from differentiating the crossing matrices) are positive. Alternatively, one must restrict attention to the massless limit, in which $\chi_u = 0$ for any value of $t$ and crossing becomes trivial.

\section{Polarisation Conventions}
\label{app:polarisation}

In this Appendix, we collect our conventions for the external states, in particular their polarisations tensors.  
Throughout we work in metric signature $(-, +, + , +)$. 

\paragraph{Spin-1:}
The polarisation tensors $\epsilon_h^\mu$ for the three helicity states of a massive vector field $A_\mu$ are given by,
\begin{align}
\epsilon_{\pm}^\mu ( \bfp_3^s ) = \frac{1}{\sqrt{2}} \left( 0,  \mp  \cos \theta_s , -i,  \pm \sin \theta_s \right)  \;, \;\;
\epsilon_{0}^\mu ( \bfp_3^s ) = \frac{1}{2m} \left(  k_s , \sqrt{s} \sin \theta_s , 0,  \sqrt{s} \cos \theta_s \right)
\label{eqn:epdef}
\end{align}
Note that $\epsilon_{h} ( \bfp ) = e^{i \pi h} \epsilon_{-h}^* ( \bfp)$, as per our CPT convention. These conventions coincide with those of \cite{deRham:2018qqo}, to allow easy comparison with those amplitudes/positivity bounds.

\paragraph{A vector example:}
For instance, consider the simple vector-scalar interaction \eqref{eqn:VS_dim6}. The inelastic amplitude $ \mathcal{A}_{+ 0 \to 0 0}$ in the $s$-channel region $s-4m^2 \geq -t \geq 0$ is given by,
\begin{align}
 \mathcal{A}_{+ 0 \to 0 0} (s,t) = - \frac{ \sqrt{s} \, \sin  \theta_s }{ 2 \sqrt{2} m }
\end{align}
using the momenta \eqref{eqn:kinematics} and polarisations \eqref{eqn:epdef}. 
Using \eqref{eqn:ths} to replace $\theta_s$ with $t$, this gives the function,
\begin{align}
\mathcal{A}_{+ 0 0 0} (s,t) =  - \lambda \frac{ \sqrt{s t u } }{ 
 \sqrt{2} m (s -4 m^2 )
 }
 \label{eqn:VS_eg_p000}
\end{align}
which can be straightforwardly continued to the whole complex $s$-plane. 
Finally, we can compare this with the value inferred from the crossing relation \eqref{eqn:helicityCrossing}, which gives,
\begin{align}
 \mathcal{A}_{+ 0 0 0} ( 4m^2 - u - t , t)  &= \sum_{h_1' h_3'} e^{i \pi ( h_1' - h_3' )} d^1_{h_1' h_1} (\chi_u) d^1_{h_3' h_3} (-\chi_u) \, \mathcal{A}_{h_1' 0 h_3' 0} ( u , t)  \nonumber \\
 &= - \lambda \frac{ \sqrt{ s t u} }{ 
 \sqrt{2} m ( u + t ) 
 }
\end{align}
in the $u$-channel region, $u-4m^2 \geq -t \geq 0$, which indeed agrees with the explicit continuation of \eqref{eqn:VS_eg_p000}.

\paragraph{Spin-1/2:}
For spinors, we adopt the conventions of \cite{Dreiner:2008tw} with spacetime signature $(-, +, +, +)$. In particular, we will explicitly write $SU(2)_L$ ($SU(2)_R$) indices $\alpha$ ($\dot \alpha$) on each Weyl spinor, which are raised and lowered as,
\begin{align}
 \lambda_\alpha = \epsilon_{\alpha \beta} \lambda^\beta \;\; , \;\;
 \lambda^\alpha = \epsilon^{\alpha \beta} \lambda_\beta \;\; ,   \;\; 
 \epsilon^{\alpha \beta} \epsilon_{\beta \gamma} = \delta^\alpha_\gamma
\end{align}
using the antisymmetric symbol $\epsilon_{\alpha \beta} = \left( \begin{array}{cc}  0 & - 1 \\ 1 & 0    \end{array} \right)$, which coincides numerically with $- \epsilon^{ \alpha \beta}$ and $\epsilon_{\dot \alpha \dot \beta}$. 

The Weyl field is quantized as $\lambda^\alpha (x) = \int_p \sum_h \left(  x^\alpha_h (\bfp) \hat{a}_h (\bfp) e^{-i p\cdot x} + y_h^\alpha (\bfp) \hat{a}_h^{\dagger} (\bfp) e^{+i p \cdot x}   \right)$, where $\hat{a}$ and $\hat{a}^\dagger$ obey a canonical anticommutation relation\footnote{
Note that it is the operators $\hat{a}_{h}, \, \hat{a}_{h}^{\dagger}$ which anti-commute, and so $x_h^{\alpha}$ and $y_h^{\alpha}$ are commuting spinors.
} and $\int_p$ is the usual integral over all on-shell, future-pointing momenta. 
The polarisations for an incoming spin-1/2 particle with mass $m$ and momentum $\mathbf{p} = k ( \sin \theta , 0, \cos \theta )$ are given by,
\begin{align}
&x^{\alpha}_{+} = \frac{m + \omega_k - k}{\sqrt{2} \sqrt{ m + \omega_k} } \left( \begin{array}{c}
\sin (\theta/2)  \\
-\cos (\theta/2 ) 
\end{array} \right)   \; ,
&\bar{y}^{\dot \alpha}_+ &= \frac{m + \omega_k + k}{\sqrt{2} \sqrt{ m + \omega_k} } \left( \begin{array}{c}
 \cos ( \theta / 2 )  \nonumber \\
 \sin ( \theta / 2 )
\end{array} \right)    \; ,  \\
&x^{\alpha}_{-} =  \frac{m + \omega_k + k}{\sqrt{2} \sqrt{ m + \omega_k} } \left( \begin{array}{c}
\cos ( \theta/2)  \\
\sin (\theta/2)
\end{array} \right) \; , 
 \quad
&\bar{y}^{\dot \alpha}_- &=  \frac{m + \omega_k - k}{\sqrt{2} \sqrt{ m + \omega_k} } \left( \begin{array}{c}
- \sin ( \theta/2)  \\
\cos (\theta/ 2)
\end{array} \right)    \; .
\end{align}
where $x_h, \bar{x}_h$ and $y_h, \bar{y}_h$ represent left-handed and right-handed chirality respectively (with $h$ labelling the helicity), and $\omega_k = \sqrt{k^2 + m^2}$ is the usual energy. The analogous outgoing states are $\bar{x}^{\dot \alpha}_{\pm}$ and $y^{\alpha}_{\pm}$, and they are related by the CPT relation, $y^\alpha_h = i e^{-i\pi h} x^\alpha_{-h}$, between an outgoing RH helicity $h$ fermion and an incoming LH helicity $-h$ fermion. 

The Dirac field is quantised as $\psi (x) = \int_p \sum_h \left(   u^h (\bfp) \hat{a}_h (\bfp) e^{-i p \cdot x} + v^h (\bfp) \hat{b}_h^{\dagger} (\bfp) e^{+i p \cdot x}         \right)$, where $\{ \hat{a}_h , \hat{a}^\dagger_h \}$ and $\{ \hat{b}_h, \hat{b}_h^\dagger \}$ obey separate anticommutation relations. 
The 4-component polarisation tensors are related to the 2-component polarisation tensors by,
\begin{align}
&\text{Incoming particle:} \;\;  u_h (\bfp) = \left( \begin{array}{c}
  x_{h \, \alpha}   \\
  \bar{y}_h^{\dot \alpha}
 \end{array} \right) \; ,     &\text{Incoming antiparticle:} \;\; \bar{v}^h ( \bfp) = \left( x^\alpha_h , \bar{y}_{h \, \dot \alpha} \right) \; , 
 \label{eqn:udef}
\end{align}
in the Weyl basis. 
The analogous outgoing states are $\bar{u}_h (\bfp) = \left( y^\alpha_h, \bar{x}_{h \, \dot \alpha} \right)$ for a particle and $v_h (\bfp) = \left( \begin{array}{c} y_{h \, \alpha}  \\ \bar{x}^{\dot \alpha}_h  \end{array} \right)$ for an antiparticle (we define the Dirac conjugate field $\bar{\Psi} = \Psi^\dagger A$), and they are related by the CPT relation, $u^h (\bfp) = i e^{-i \pi h} \gamma_5 v^{-h} (\bfp)$, and also by charge conjugation, $v^h (\bfp) = C \bar{u}^h (\bfp)^T$, where,
\begin{align}
\gamma^\mu = \left(  \begin{array}{c c}
0 & \sigma^\mu_{\alpha \dot \beta}  \\
\bar{\sigma}^{\mu \; \dot \alpha  \beta} & 0
\end{array}
 \right) \;, \qquad
 \gamma_5 = \left( \begin{array}{c c}
 - \delta_\alpha^{\; \beta} & 0 \\
 0 &  \delta^{\dot \alpha}_{\; \dot \beta}
 \end{array} \right) \qquad 
A = \left( \begin{array}{c c}
0 & \delta^{\dot \alpha}_{\; \dot \beta}  \\
\delta_\alpha^{\; \beta} & 0 
\end{array} \right) \; , \qquad
C = \left( \begin{array}{c c}
\epsilon_{\alpha \beta} & 0 \\
0 & \epsilon^{\dot \alpha \dot \beta}
\end{array} \right) \; . 
\end{align}
in the Weyl basis. $\sigma^\mu_{\alpha \dot \beta}$ coincides numerically with the usual $2 \times 2$ Pauli matrices, $\left( \mathbbm{1} , \sigma_x, \sigma_y, \sigma_z \right)$, while $\bar{\sigma}^{\mu \; \dot \alpha \beta}$ coincides numerically with $\left( \mathbbm{1} , -\sigma_x, -\sigma_y, -\sigma_z \right)$. These tensors satisfy a variety of useful identities (see \cite{Dreiner:2008tw}), and in particular we will make use of,
\begin{align}
\eta_{\mu\nu} \sigma^\mu_{\alpha \dot \alpha} \sigma^\nu_{\beta \dot \beta} = - 2 \epsilon_{\alpha \beta} \epsilon_{\dot \alpha \dot \beta} \;\; , \;\; \eta^{\mu\nu} \bar{\sigma}^{\dot \alpha \alpha}_\mu \bar{\sigma}^{\dot \beta  \beta}_\nu = - 2 \epsilon^{\alpha \beta} \epsilon^{\dot \alpha \dot \beta} \;\; , \;\;  \sigma^\mu_{\alpha \dot \alpha} \bar{\sigma}^{\dot \beta \beta}_\mu = - 2 \delta_{\alpha}^{\beta} \delta_{\dot \alpha}^{\dot \beta} \; .
\label{eqn:sigma_identities}
\end{align}

\paragraph{A spinor example:}
Consider the simple quartic interaction $\tfrac{1}{2} \left( \bar{\psi} \psi \right)^2$. 
At tree-level, the on-shell amplitude for the $\psi \psi \to \psi \psi$ process is,
\begin{align}
\mathcal{A}_{h_1 h_2 \to h_3 h_4}^{\psi \psi \to \psi \psi}  &= \langle 0 | \hat{a}_{h_3} (  \mathbf{p}_3^s ) \hat{a}_{h_4} (  \mathbf{p}_4^s ) \left( \bar{\psi} \psi \right)^2 \hat{a}_{h_2}^\dagger \left(  \mathbf{p}_2^s  \right) \hat{a}_{h_1}^\dagger (  \mathbf{p}_1^s ) | 0 \rangle   \label{eqn:FFeg_1}  \\
&= + \left(  \bar{u}_{h_3} (  \mathbf{p}_3^s )  u_{h_1} ( \mathbf{p}_1^s ) \right)  \left( \bar{u}_{h_4} (  \mathbf{p}_4^s ) u_ {h_2} \left(  \mathbf{p}_2^s  \right) \right) 
  -  \left(  \bar{u}_{h_4} (  \mathbf{p}_4^s )  u_{h_1} ( \mathbf{p}_1^s ) \right)  \left( \bar{u}_{h_3} (  \mathbf{p}_3^s ) u_ {h_2} \left(  \mathbf{p}_2^s  \right) \right) \, .    \nonumber 
\end{align}
For instance, using \eqref{eqn:kinematics} and \eqref{eqn:udef} for the momenta and polarisations, this gives $\mathcal{A}_{+- \to +-}^{\psi \psi \to \psi \psi} = - 4 k^2 \cos^2 \left( \theta_s/ 2\right)$. 
The analytic continuation $\mathcal{A}_{h_1 h_2 h_3 h_4} (s,t)$ to complex values of $s$ is defined such that it is analytic off the real axis and coincides with \eqref{eqn:FFeg_1} on the real axis (approached from above)---for instance, using \eqref{eqn:ths} gives  $\mathcal{A}_{+-+-}^{\psi \psi \to \psi \psi} (s,t) = u$ for all real $s - 4m^2 \geq -t \geq 0$, which is straightforwardly continued to  $\mathcal{A}_{+-+-}^{\psi \psi \to \psi \psi} (s,t) = u$ in the whole complex $s$-plane. 

The corresponding $u$-channel amplitude for the process $\psi \bar{\psi} \to \psi \bar{\psi}$ is given by,
\begin{align}
\mathcal{A}_{h_1 \bar{h}_4 \to h_3 \bar{h}_2}^{\psi \bar{\psi} \to \psi \bar{\psi} }  &= \langle 0 | \hat{a}_{h_3} (  \mathbf{p}_3^u ) \hat{b}_{\bar{h}_2} (  \mathbf{p}_2^u ) \left( \bar{\psi} \psi \right)^2 \hat{b}_{\bar{h}_4}^\dagger \left(  \mathbf{p}_4^u  \right) \hat{a}_{h_1}^\dagger (  \mathbf{p}_1^u ) | 0 \rangle    \\
&= - \left(  \bar{u}_{h_3} (  \mathbf{p}_3^u )  u_{h_1} ( \mathbf{p}_1^u ) \right)  \left( \bar{v}_{\bar{h}_4} (  \mathbf{p}_4^u ) v_ {\bar{h}_2} \left(  \mathbf{p}_2^u  \right) \right) 
  + \left(  \bar{v}_{\bar{h}_4} (  \mathbf{p}_4^u )  u_{h_1} ( \mathbf{p}_1^u ) \right)  \left( \bar{u}_{h_3} (  \mathbf{p}_3^u ) v_ {\bar{h}_2} \left(  \mathbf{p}_2^u  \right) \right) \, ,    \nonumber 
\end{align}
which can be analogously continued into the complex plane to define the complex function $\mathcal{A}_{h_1 \bar{h}_4 h_3 \bar{h}_2}^{\psi \bar{\psi} \to \psi \bar{\psi} } (u ,t)$. 
Explicitly, the complete list of amplitudes is given by, 
\begin{align}
\mathcal{A}_{+ + + +}^{\psi \psi \to \psi \psi } (s ,t) &= 4m^2  \; , 
&\mathcal{A}_{+ - + -}^{\psi \bar{\psi} \to \psi \bar{\psi} } (u ,t) &= 4m^2 - \frac{4m^2 t}{s+t} \; , \nonumber \\
\mathcal{A}_{+ + - -}^{\psi \psi \to \psi \psi } (s ,t) &= s  \; , 
&\mathcal{A}_{+ + - -}^{\psi \bar{\psi} \to \psi \bar{\psi} } (u ,t) &= s + \frac{ 4 m^2 t}{ s + t } \; , \nonumber    \\
\mathcal{A}_{+ - - +}^{\psi \psi \to \psi \psi } (s ,t) &= t \; , 
&\mathcal{A}_{+ - - +}^{\psi \bar{\psi} \to \psi \bar{\psi} } (u ,t) &= t - \frac{4m^2 t}{s+t} \; , \label{eqn:FFeg_1_allA} \\
\mathcal{A}_{+ - + -}^{\psi \psi \to \psi \psi } (s ,t) &= u  \; , 
&\mathcal{A}_{+ + + +}^{\psi \bar{\psi} \to \psi \bar{\psi} } (u ,t) &= u - \frac{4 m^2 t }{ s + t }  \; ,
 \nonumber \\
\mathcal{A}_{+ + + -}^{\psi \psi \to \psi \psi } (s ,t) &= 0  \; , 
&\mathcal{A}_{+ + + -}^{\psi \bar{\psi} \to \psi \bar{\psi} } (u ,t) &= \frac{ 2 m \sqrt{ s t u} }{ s + t }  \; , \nonumber 
\end{align} 
together with the relations $\mathcal{A}_{-h_1, -h_2, -h_3, -h_4} = (-1)^{ h_s^{\rm in}  - h_s^{\rm out} } \mathcal{A}_{h_1 h_2 h_3 h_4} = \mathcal{A}_{h_3 h_4 h_1 h_2}$ which follow from the parity and time reversal invariance of the interaction (see e.g. Appendix E of \cite{deRham:2017zjm} for a simple derivation), and $\mathcal{A}_{h_1 h_2 h_3 h_4} = \mathcal{A}_{h_2 h_1 h_4 h_3}$ since the particles are identical. 

Note that the $u$-channel function $\mathcal{A}_{h_1 \bar{h}_4 h_3 \bar{h}_2}^{\psi \bar{\psi} \to \psi \bar{\psi} } (u,t)$ is not independent of the $s$-channel $\mathcal{A}_{h_1 h_2 h_3 h_4}^{\psi \psi \to \psi \psi } (s,t)$, and indeed the two are related by the crossing equation~\eqref{eqn:helicityCrossing}, which can be checked explicitly using \eqref{eqn:FFeg_1_allA}. In the forward or massless limits (i.e. $4m^2 t \to 0$), this crossing relation becomes the trivial \eqref{eqn:crossingForward}, which in this case corresponds to simply $\mathcal{A}_{h_1 h_2 h_3 h_4}^{\psi \psi \to \psi \psi } (s,t) = \mathcal{A}_{h_1 \bar{h}_4 h_3 \bar{h}_2}^{\psi \bar{\psi} \to \psi \bar{\psi} } (u,t)$.

The regulated elastic $\hat{A}_{h_u}$ amplitudes are given by,
\begin{align}
\hat{A}^{\psi \psi \to \psi \psi}_{\pm 1} (s,t) &= 4m^2 s (s-4m^2)  \; ,   &\hat{A}^{\psi \bar{\psi} \to \psi \bar{\psi} }_{\pm 1} (s,t) &=  s^3 + 4 m^2 s (t - s) \; ,   \nonumber  \\ 
\hat{A}^{\psi \psi \to \psi \psi}_{0} (s,t) &= 4m^2 - s   \; , &\hat{A}^{\psi \bar{\psi} \to \psi \bar{\psi} }_{0} (s,t) &= 4m^2 \; . 
\end{align}
Note that our positivity bounds do not place any constraint on this dimension-6 operator unless the UV amplitude converges fast enough for the $\mathcal{A}(s,t)$ dispersion relation to converge with zero subtractions (i.e. $N_{\rm UV} = 0$), which is stronger than both the Froissart and the super-Froissart conditions considered in the main text.

\paragraph{Another spinor example:}
Finally, consider the quartic interaction $\tfrac{1}{2} \left( \bar{\psi} \gamma_\mu \psi \right)^2$. 
At tree-level, the on-shell amplitudes for the $s$- and $u$-channel processes are,
\begin{align}
\mathcal{A}_{h_1 h_2 \to h_3 h_4}^{\psi \psi \to \psi \psi} 
&= + \left(  \bar{u}_{h_3} (  \mathbf{p}_3^s ) \gamma_\mu u_{h_1} ( \mathbf{p}_1^s ) \right)  \left( \bar{u}_{h_4} (  \mathbf{p}_4^s ) \gamma^\mu u_ {h_2} \left(  \mathbf{p}_2^s  \right) \right) 
  \nonumber \\
  &\qquad-  \left(  \bar{u}_{h_4} (  \mathbf{p}_4^s ) \gamma_\mu u_{h_1} ( \mathbf{p}_1^s ) \right)  \left( \bar{u}_{h_3} (  \mathbf{p}_3^s ) \gamma^\mu u_ {h_2} \left(  \mathbf{p}_2^s  \right) \right)  \label{eqn:FFeg_2}  \\
 \mathcal{A}_{h_1 \bar{h}_4 \to h_3 \bar{h}_2}^{\psi \bar{\psi} \to \psi \bar{\psi} }  &=  - \left(  \bar{u}_{h_3} (  \mathbf{p}_3^u )  \gamma_\mu u_{h_1} ( \mathbf{p}_1^u ) \right)  \left( \bar{v}_{\bar{h}_4} (  \mathbf{p}_4^u ) \gamma^\mu v_ {\bar{h}_2} \left(  \mathbf{p}_2^u  \right) \right)  \nonumber\\
 &\qquad
  + \left(  \bar{v}_{\bar{h}_4} (  \mathbf{p}_4^u ) \gamma_\mu  u_{h_1} ( \mathbf{p}_1^u ) \right)  \left( \bar{u}_{h_3} (  \mathbf{p}_3^u ) \gamma^\mu  v_ {\bar{h}_2} \left(  \mathbf{p}_2^u  \right) \right) \, ,    \nonumber 
\end{align}
Evaluating these for each choice of helicity, and then using \eqref{eqn:ths} to write the result in terms of $s$ and $t$, gives,
\begin{align}
\mathcal{A}_{+ + + +}^{\psi \psi \to \psi \psi } (s ,t) &= -4 s + 12 m^2   \; , 
&\mathcal{A}_{+ - + -}^{\psi \bar{\psi} \to \psi \bar{\psi} } (u ,t) &= -4 s + 12 m^2 - \frac{12 m^2 t}{ s + t }  \; , \nonumber \\
\mathcal{A}_{+ + - -}^{\psi \psi \to \psi \psi } (s ,t) &= - 4 m^2   \; , 
&\mathcal{A}_{+ + - -}^{\psi \bar{\psi} \to \psi \bar{\psi} } (u ,t) &= - 4m^2 + \frac{12 m^2 t}{s+t}  \; , \nonumber    \\
\mathcal{A}_{+ - - +}^{\psi \psi \to \psi \psi } (s ,t) &= 2 t \; , 
&\mathcal{A}_{+ - - +}^{\psi \bar{\psi} \to \psi \bar{\psi} } (u ,t) &= 2 t - \frac{ 12 m^2 t }{ s + t }  \; , \label{eqn:FFeg_2_allA} \\
\mathcal{A}_{+ - + -}^{\psi \psi \to \psi \psi } (s ,t) &=  2u  \; , 
&\mathcal{A}_{+ + + +}^{\psi \bar{\psi} \to \psi \bar{\psi} } (u ,t) &= 2u - \frac{ 12 m^2 t }{ s + t }  \; ,
 \nonumber \\
\mathcal{A}_{+ + + -}^{\psi \psi \to \psi \psi } (s ,t) &= 0  \; , 
&\mathcal{A}_{+ + + -}^{\psi \bar{\psi} \to \psi \bar{\psi} } (u ,t) &= \frac{ 6 m \sqrt{ s t u} }{ s + t }  \; , \nonumber 
\end{align} 
where again the other amplitudes follow from parity, time reversal and particle exchange. Each of these functions is straightforwardly continued from the physical $s$- and $u$-channel regions to the entire complex plane, and again we find that $\mathcal{A}_{h_1 h_2 h_3 h_4}^{\psi \psi \to \psi \psi } (s,t)$ and $\mathcal{A}_{h_1 \bar{h}_4 h_3 \bar{h}_2}^{\psi \bar{\psi} \to \psi \bar{\psi} } (u,t)$ are related by the crossing equation~\eqref{eqn:helicityCrossing}. 

The regulated elastic $\hat{A}_{h_u}$ amplitudes are given by,
\begin{align}
\hat{A}^{\psi \psi \to \psi \psi}_{\pm 1} (s,t) &= - 4 s (s-4m^2) (s - 3 m^2)  \; ,   &\hat{A}^{\psi \bar{\psi} \to \psi \bar{\psi} }_{\pm 1} (s,t) &= 2 s^3  + 4 m^2 s \left( 3 t - s \right)    \nonumber  \\ 
\hat{A}^{\psi \psi \to \psi \psi}_{0} (s,t) &= - 2 (s - 4m^2 )   \; , &\hat{A}^{\psi \bar{\psi} \to \psi \bar{\psi} }_{0} (s,t) &=  4 s - 4m^2
\end{align}
Note that our positivity bounds also do not place any constraint on this particular dimension-6 operator unless the dispersion relation converges with zero subtractions (i.e. $N_{\rm UV} = 0$), which is stronger than both the Froissart and the super-Froissart conditions considered in the main text.

\paragraph{Massless Limit:}
In the massless limit, $m \to 0$ ($\omega_k \to k$), there are only two non-zero polarisations, one for each helicity (which now coincides with the chirality). These are often denoted using the angled- and square-bracket spinor helicity variables,
\begin{align}
| n \rangle^{\dot \alpha} :=  \bar{x}_-^{\dot \alpha} (p_n)  =  \bar{y}_{+}^{\dot \alpha} (p_n )   \;\; \text{and} \;\;  [ n |^\alpha :=  y_+^\alpha  (p_n) = x_{-}^\alpha (p_n ) \;\;\; \text{when } m =0 \; , 
\label{eqn:spinorhelicitydef}
\end{align}
whose indices are lowered via $\langle p |_{\dot \alpha} = \epsilon_{\dot \alpha \dot \beta} | p \rangle_{\dot \beta}$ and $| p ]_\alpha = \epsilon_{\alpha \beta} [ p |^\beta$, and which obey the familiar relation $| n \rangle^{\dot \alpha} [ n |^\beta =  -p_n^\mu \sigma_\mu^{\dot \alpha \beta} $, where $p_n$ is the momentum of particle $n$. See \cite{Elvang:2013cua} for a review.

Since crossing is trivial in this limit, it is straightforward to relate the different channels, and in particular it is conventional to consider the particles as either all incoming or all outgoing. 
For instance, another way to compute $\mathcal{A}^{\psi \bar{\psi} \to \psi \bar{\psi}}_{h_1 \bar{h}_2 h_3 \bar{h}_4}$ in the massless limit is to first consider the amplitude for all particles incoming\footnote{
Note that \eqref{eqn:FFeg_2_spinorhelicity} is a purely off-shell expression, since there are no (real) on-shell 4-momenta which can satisfy momentum conservation ($\sum_n p_n = 0$) when all incoming. 
},  
\begin{align}
\mathcal{A}^{ \psi \bar{\psi} \bar{\psi} \psi \to \varnothing}_{h_1 \bar{h}_2 \bar{h}_3 h_4} ( p_1, p_2, p_3, p_4 )  &=  \langle 0 |   \left( \bar{\psi} \gamma_\mu \psi \right)^2  \hat{a}_4^\dagger \,  \hat{b}_3^\dagger \, \hat{b}^\dagger_2 \, \hat{a}_{1}^\dagger \, | 0 \rangle  \\
&= \left( \bar{v}_3 \gamma_\mu u_1 \right) \left( \bar{v}_2 \gamma^\mu u_4 \right) - \left( \bar{v}_2 \gamma_\mu u_1 \right) \left( \bar{v}_3 \gamma^\mu u_4 \right) \; ,
\end{align}
or in terms of the spinor helicities \eqref{eqn:spinorhelicitydef},
\begin{align}
\mathcal{A}^{ \psi \bar{\psi} \bar{\psi} \psi \to \varnothing}_{+ - - +} (p_1, p_2, p_3, p_4) &=   - 4\,  \langle 1 4 \rangle [ 2 3 ] \; ,
\label{eqn:FFeg_2_spinorhelicity}
\end{align}
where we have used the identities \eqref{eqn:sigma_identities}. 
Then by using crossing,
\begin{align}
 \mathcal{A}^{ \psi \bar{\psi} \bar{\psi} \psi \to \varnothing}_{h_1 \bar{h}_2 \bar{h}_3 h_4} (p_1, p_2, -p_3, -p_4) = \mathcal{A}^{\psi \bar{\psi} \to \psi \bar{\psi}}_{h_1 \bar{h}_2 h_3 \bar{h}_4} ( p_1, p_2, p_3, p_4 )
 \label{eqn:crossing_ppmm}
\end{align}
together with the analytic continuation of the spinor helicities, $ | - p \rangle = - | p \rangle$ and $| - p ] = + | p \rangle $, the simple expression \eqref{eqn:FFeg_2_spinorhelicity} implies that,
 \begin{align}
\mathcal{A}^{\psi \bar{\psi} \to \bar{\psi} \psi}_{+ - + -} ( p_1, p_2, p_3, p_4 )
=  + 4 \langle 1 4 \rangle [ 2 3 ]  
\end{align}
which for the kinematics \eqref{eqn:kinematics_u} gives,
\begin{align}
\mathcal{A}^{\psi \bar{\psi} \to \bar{\psi} \psi}_{+ - + -} ( s , t ) =  +4s + 4t
\label{eqn:-4u}
\end{align}
in perfect agreement with the massless limit of \eqref{eqn:FFeg_2_allA}. 

However, note that care must be taken when applying the crossing relation from the all-incoming process to the physical $2 \to 2$ process---for instance, had we instead crossed particles 2 and 4, the correct relation is, 
\begin{align}
 \mathcal{A}^{ \psi \bar{\psi} \bar{\psi} \psi \to \varnothing}_{h_1 \bar{h}_2 \bar{h}_3 h_4} (p_1, -p_2, p_3, -p_4) =  - \mathcal{A}^{\psi \bar{\psi} \to \psi \bar{\psi} }_{h_1 \bar{h}_3 h_2 \bar{h}_4} ( p_1, p_3, p_2, p_4 ) \;  ,
 \label{eqn:crossing_pmpm}
\end{align}
which differs from \eqref{eqn:crossing_ppmm} by an overall minus sign\footnote{
The relative minus sign is due to the Fermi statistics of the particles, and is most easily seen by considering the permutation of particles 2 and 3, $\mathcal{A}^{\psi \bar{\psi} \bar{\psi} \psi \to \varnothing}_{h_1 \bar{h}_2 \bar{h}_3 h_4} (p_1, p_2, p_3, p_4) = - \mathcal{A}^{\psi \bar{\psi} \bar{\psi} \psi \to \varnothing}_{h_1 \bar{h}_3 \bar{h}_2 h_4} ( p_1 , p_3, p_2 , p_4)$.
}. 
As an example, suppose that we take $p_2 = - p_1$ and $p_4 = -p_3$ in order to set $s=0$,
\begin{align}
\mathcal{A}^{ \psi \bar{\psi} \bar{\psi} \psi \to \varnothing}_{+ - - +} (p_1, -p_1,  p_3, -p_3) &=  + 4\,  \langle 1 3 \rangle [ 1 3 ] = - 4 t  \; ,
\end{align}
then this corresponds via the crossing relation \eqref{eqn:crossing_pmpm} to,
\begin{align}
- \mathcal{A}^{ \psi  \bar{\psi} \to \psi \bar{\psi} }_{+ - + -} (s,t) |_{s=0}  = - 4 t \; ,
\label{eqn:-4t}
\end{align}
which indeed reproduces the massless limit\footnote{
Note that the $m \to 0$ limit is performed first to define the massless amplitude, and then is followed by the kinematic limit $s \to 0$. 
} of \eqref{eqn:FFeg_2_allA}. 
This overall sign in \eqref{eqn:crossing_pmpm} is crucial for correctly applying the positivity bounds, since without it one might conclude from \eqref{eqn:-4u} and \eqref{eqn:-4t} that $\partial_t \mathcal{A}_{+-+-}^{\psi \bar{\psi} \to \psi \bar{\psi}} (s,t) |_{s=0}$ and $\partial_s \mathcal{A}_{+-+-}^{\psi \bar{\psi} \to \psi \bar{\psi}} (s,t)|_{t=0}$ have opposite signs, when it is clear from \eqref{eqn:FFeg_2_allA} that they in fact have the same sign and so cancel out in \eqref{eqn:RR}.

\section{Scattering Unequal Masses}
\label{app:unequal}

For algebraic simplicity, we have focussed in the main text on scattering processes in which the four external particles have the same mass, $m$. This restriction is by no means necessary, and in this Appendix we derive the analogous positivity bounds for the scattering of unequal mass particles. 

\paragraph{Kinematics:}
Consider the elastic scattering process between two particles with masses $m_1, m_2$ and spins $S_1, S_2$. We label the particles so that $m_1 \geq m_2$, and define the positive difference,
\begin{align}
\Delta = m_1^2 - m_2^2 \; ,
\end{align}
along with the following convenient pair of analytic functions,
\begin{align}
\mathcal{S} = (s - (m_1 - m_2)^2 )(s - (m_1 + m_2 )^2 ) \; , \nonumber \\
\mathcal{U} = (u - (m_1 - m_2 )^2 ) ( u - (m_1 + m_2 )^2 ) \; .
\end{align}
The momenta of the particles in the $s$-channel centre-of-mass frame remains \eqref{eqn:kinematics}, but now with a scattering angle $\theta_s$ given by,
\begin{align}
 \cos \frac{\theta_s}{2} = \frac{ \sqrt{ - s u + \Delta^2 } }{ \sqrt{\mathcal{S}} } \;\; , \;\;\;\; \sin \frac{\theta_s}{2} = \frac{\sqrt{-s t}}{ \sqrt{\mathcal{S}} } \; .
 \label{eqn:ths_general}
\end{align}

Repeating the steps in Section~\ref{sec:unitarity} leads to the same partial wave expansion remains~\eqref{eqn:Apw}, but with $\theta_s$ now given by \eqref{eqn:ths_general} in place of \eqref{eqn:ths}. 
Consequently, $t$- and $\theta_s$-derivatives are now related by the factor $\mathcal{S} \partial_t = 2 s  \partial/\partial \cos \theta_s$, and so \eqref{eqn:unit_As} becomes,
\begin{align}
\frac{\mathcal{S}}{s} \partial_t \, \text{Abs}_s \, \mathcal{A}_{h_1 h_2 h_1 h_2} (s,t) |_{t=0} \geq |h_s| \, \text{Abs}_s \, \mathcal{A}_{h_1 h_2 h_1 h_2} (s, 0 ) \; .
\label{eqn:unit_As_m1m2}
\end{align} 

Similarly in the $u$-channel, the centre-of-mass momenta~\eqref{eqn:kinematics_u} are now given by the scattering angle,
\begin{align}
 \cos \frac{\theta_u}{2} = \frac{ \sqrt{ - s u + \Delta^2 } }{ \sqrt{\mathcal{U}} } \;\; , \;\;\;\; \sin \frac{\theta_u}{2} = \frac{\sqrt{-u t}}{ \sqrt{\mathcal{U}} } \; ,
\end{align}
in place of \eqref{eqn:thu}, and the partial wave expansion is again~\eqref{eqn:Au_pw}.

\paragraph{Crossing:}
When the masses are unequal, the rest-frame rotation \eqref{eqn:restFrameRotation} required to go from $s$- to $u$-channel kinematics acts differently on each particle: in particular the angles $\chi_a$ are no longer given by \eqref{eqn:chiu}, but rather by,
\begin{align}
\label{eq:chi1}
 \cos \chi_1 = \frac{ - (s + \Delta) (u + \Delta ) + 4 m_1^2 \Delta }{ \sqrt{  \mathcal{S} \mathcal{U} } } \,, & \quad &
 \sin \chi_1 = \frac{+ 2 m_1 \sqrt{ - t \Psi  }  }{ \sqrt{ \mathcal{S} \mathcal{U} } }  ,
 \\
 \label{eq:chi2}
 \cos \chi_2 =  \frac{ -(s - \Delta) (u - \Delta ) - 4 m_2^2 \Delta }{ \sqrt{  \mathcal{S} \mathcal{U} } } \,,
& \quad &
 \sin \chi_2 = \frac{ - 2 m_2 \sqrt{ - t \Psi  }  }{ \sqrt{ \mathcal{S} \mathcal{U} } }  ,
 \end{align}
and $\chi_3 = -\chi_1$, $\chi_4 = -\chi_2$. 
This clearly reduces to \eqref{eqn:AngleChoice} when the mass difference $\Delta \to 0$.  

The first $t$-derivative of the crossing matrices in \eqref{eqn:Across_pw} is now,
\begin{align}
&\frac{\mathcal{U}}{u} \partial_t \, \text{Abs}_u \, \mathcal{A}_{h_1 h_2 h_1 h_2} ( 2m_1^2 + 2m_2^2  -u - t, t ) |_{t=0} \nonumber \\
&= \sum_{J_u = |h_u|}^{\infty} \langle J_u \; h_u | \hat{J}_y^2 | J_u \; h_u \rangle | \langle T_u | h_1 \bar{h}_2 \rangle |^2 +  \frac{2 (m_1+m_2)^2 }{u}   \langle \hat{s}_y^2 \rangle \; , 
\end{align}
where $\langle \hat{s}_y^2 \rangle$ is given by \eqref{eqn:Sy2} with the operator $\hat{S}_y | h_1 h_2 \rangle = \left( \hat{S}_y^{(1)} + \hat{S}_y^{(2)}   \right)|h_1 h_2 \rangle$ replaced by,
\begin{align}
\hat{s}_y | h_1 h_2 \rangle = \left(  \frac{m_1^2}{(m_1+m_2)^2 } \hat{S}_y^{(1)} + \frac{m_2^2}{(m_1+m_2)^2 } \hat{S}_y^{(2)}   \right)|h_1 h_2 \rangle \; .
\end{align}
Since $\hat{s}_y$ is Hermitian, the helicity sum defined in \eqref{eqn:AhuDef} ensures that,
\begin{align}
\sum_{ \substack{ h_1 \bar{h}_2 \\ h_1 - \bar{h}_2 = h_u }} \langle \hat{s}_y \rangle \geq 0 \; , 
\label{eqn:sy2_pos}
\end{align} 
and therefore,
\begin{align}
\frac{\mathcal{U}}{u} \partial_t \, \text{Abs}_u \,  \mathcal{A}_{h_u} (2m_1^2 + 2m_2^2  -u -t, t) |_{t=0} \geq |h_u| \, \text{Abs}_u \, \mathcal{A}_{h_u} ( 2m_1^2 + 2m_2^2 - u, 0 ) \; ,
\end{align} 
which is the crossing image of \eqref{eqn:unit_As_m1m2}. 
It is perhaps worth commenting here that, without the succinct operator notation in \eqref{eqn:Sy2}, the positivity of \eqref{eqn:sy2_pos} would not have been at all obvious, since writing these sums out explicitly in terms of e.g. Wigner $d$ matrices leads to a lengthy expression with many terms.

\paragraph{Analyticity:}
The amplitude $\mathcal{A}_{h_1 h_2 h_3 h_4} (s,t)$ contains unphysical kinematic singularities arising from the factors of $\cos \theta_s/2$ and $\sin \theta_s/2$ used to define the polarisation tensors.
These kinematic singularities were studied in detail in \cite{Cohen-Tannoudji:1968kvr}, where it was shown that for elastic processes in which the helicities are preserved, the \emph{regulated} amplitude,
\begin{align}
 \hat{\mathcal{A}}_{h_1 h_2 h_1 h_2} (s, t) = \frac{\mathcal{S}^{S_1 + S_2}  }{ \left( - s u + \Delta^2 \right)^{|  h_s |} } \mathcal{A}_{h_1 h_2 h_1 h_2} (s,t) \; , 
\end{align}
is free from any unphysical kinematic singularity, where $h_s = h_1 - h_2$ as in the main text. The only non-analyticities of $ \hat{\mathcal{A}}_{h_1 h_2 h_1 h_2} (s, t)$ in the complex $s$-plane are those required by unitarity and crossing, namely,
\begin{align}
&\text{Poles:}  \;\; s = m_1^2 \; ,  \;\;  m_2^2 \; , \;\;   2 m_1^2 + m_2^2 - t \; , \;\; m_1^2 + 2 m_2^2 -t \; ,     \nonumber \\
&\text{Branch Cuts:}  \;\; s \geq (m_1 + m_2 )^2   \; \;\; \text{and} \;\;\; s \leq (m_1 - m_2 )^2 - t  \; , 
\end{align}
as well as for the masses and thresholds of any other fields which couple to the external particles. 

As in section~\ref{sec:positivity}, the $s$-channel branch cut of $\hat{A}_{h_1 h_2 h_1 h_2}$ has positive $t$-derivative, 
\begin{align}
\frac{\mathcal{S}}{s} \partial_t \text{Abs}_s \, \hat{\mathcal{A}}_{h_1 h_2 h_1 h_2} (s,t) |_{t=0} &= \mathcal{S}^{S_1 + S_2 - |h_s|}  \left( 
 \frac{\mathcal{S}}{s}  \partial_t - | h_s|  
 \right) \text{Abs}_s \, \mathcal{A}_{h_1 h_2 h_1 h_2} (s,t) |_{t=0} \nonumber \\
&\geq 0 \; . 
\label{eqn:Is_m1m2}
\end{align}
Similarly, the $u$-channel branch cut is bounded by,
\begin{align}
&\frac{\mathcal{U}}{u} \partial_t \text{Abs}_u \, \hat{\mathcal{A}}_{h_u} ( 2m_1^2 + 2m_2^2 -u-t,t) |_{t=0} \nonumber\\
 &\geq \mathcal{U}^{S_1 + S_2 - |h_s|_{\rm min}}  \left( 
|h_u| - |h_s|_{\rm min} +  \left( S_1 + S_2 \right) \left(  1 + \frac{u - 2m_1^2 - 2m_2^2 }{u}  \right)
 \right) \nonumber \\ 
 &\quad\qquad\qquad\quad\quad \times \text{Abs}_s \, \mathcal{A}_{h_u} ( 2m_1^2 + 2m_2^2 - u , 0)  \;  .
 \label{eqn:Iu_m1m2}
\end{align}
The dispersion relation is then given by \eqref{eqn:dt_disp}, and the branch cuts $I_s$ and $I_u$ are positive thanks to \eqref{eqn:Is_m1m2} and \eqref{eqn:Iu_m1m2}, completing the proof of the positivity bound for generic particle masses $m_1 \neq m_2$.

\bibliographystyle{JHEP}
\bibliography{Dim6_Bounds}

\end{document}